\documentclass[pre,twocolumn,superscriptaddress, aps]{revtex4-2}
\usepackage[utf8]{inputenc}

\usepackage[separate-uncertainty = true,multi-part-units=single]{siunitx}

\usepackage[hidelinks]{hyperref}

\usepackage{changes}
\usepackage{graphicx}
\usepackage{amsmath}
\usepackage{amssymb}
\usepackage{todonotes}

\usepackage{mhchem}
\usepackage{chemformula}
\setchemformula{kroeger-vink}
\DeclareSIUnit\angstrom{\text{Å}}

\presetkeys%
    {todonotes}%
    {inline,backgroundcolor=white}{}
\makeatletter

\makeatletter
\renewcommand{\todo}[2][]{\@bsphack\@todo[#1]{\textcolor{black!70}{#2}}\@esphack\ignorespaces}
\makeatother

\begin{document}

\title{Structure and thermodynamics of defects in Na-feldspar from a neural network potential}

\author{Alexander Gorfer}
\affiliation{Faculty of Physics, University of Vienna, Boltzmanngasse 5, 1090, Vienna, Austria}
\affiliation{Department of Lithospheric Research, University of Vienna, Josef-Holaubuek-Platz 2, 1090, Vienna, Austria}
\affiliation{Vienna Doctoral School in Physics, University of Vienna, Boltzmanngasse 5, 1090, Vienna, Austria}
\author{Rainer Abart}
\affiliation{Department of Lithospheric Research, University of Vienna, Josef-Holaubuek-Platz 2, 1090, Vienna, Austria}
\author{Christoph Dellago}
\affiliation{Faculty of Physics, University of Vienna, Boltzmanngasse 5, 1090, Vienna, Austria}

\begin{abstract}
The diffusive phase transformations occurring in feldspar, a common mineral in the crust of the Earth, are essential for reconstructing the thermal histories of magmatic and metamorphic rocks. Due to the long timescales over which these transformations proceed, the mechanism responsible for sodium diffusion and its possible anisotropy has remained a topic of debate. To elucidate this defect-controlled process, we have developed a Neural Network Potential (NNP) trained on first-principle calculations of Na-feldspar (Albite) and its charged defects. This new force field reproduces various experimentally known properties of feldspar, including its lattice parameters, elastic constants as well as heat capacity and DFT-calculated defect formation energies. A new type of dumbbell interstitial defect is found to be most favorable and its free energy of formation at finite temperature is calculated using thermodynamic integration. The necessity of including electrostatic corrections before training an NNP is demonstrated by predicting more consistent defect formation energies.
\end{abstract}
\maketitle

\section{Introduction}
Feldspar is the most abundant mineral in the Earth's crust and an important ingredient for ceramics \cite{Dondi2018}. It forms a solid-solution between a sodium (NaAlSi$_3$O$_8$ - albite), calcium (CaAl$_2$Si$_2$O$_8$ - anorthite) and a potassium (KAlSi$_3$O$_8$ - K-feldspar) end-member component.  %
The most prominent feldspar groups are represented by the plagioclase (NaAlSi$_3$O$_8$ - CaAl$_2$Si$_2$O$_8$) and the alkali feldspar (NaAlSi$_3$O$_8$ - KAlSi$_3$O$_8$) solid-solution series. At high temperatures typical for magmatic and metamorphic environments, both solid-solution series show complete miscibility. Towards lower temperatures miscibility gaps open, and feldspars of intermediate composition tend to exsolve, typically producing lamellar intergrowth of feldspars with different compositions. The exsolution microstructures are of particular interest for reconstructing the thermal history of magmatic and metamorphic rocks \cite{Parsons2015, Abart2009}. In addition, elastic strain associated with exsolution microstrutures may lead to pseudocleavage, which substantially enhances the ice nucleation activity of exsolved feldspar making the corresponding aerosol particles important players in the glaciation of clouds \cite{Kiselev2017}. 

Exsolution of an initially homogeneous feldspar implies segregation of the different cations on the alkali and alkali earth sub-lattices, which occurs by intracrystalline diffusion. Due to the long timescales over which diffusive phase transformations in feldspar proceed, the underlying diffusion mechanisms and their possible anisotropy still hold unresolved questions, such as the recently discussed systematic differences between experimentally determined Na-K interdiffusion coefficients and those theoretically predicted from measured tracer diffusion coefficients \cite{Petrishcheva2014, Petrishcheva2020, ,Hergemoeller2017Potassium, Wilangowski2015a, ElMaanaoui2016}. 

On the computational side, the diffusion of alkali defects in feldspar has been studied with kinetic Monte Carlo simulations \cite{Wilangowski2015b, Wilangowski2015, Wilangowski2017} using empirical rate constants. Interatomic potentials capturing all constituent elements of the material and its defects are needed for studying the microscopic mechanism of diffusion without referring to laboratory experiments. It has been demonstrated that classical force-fields are capable of reproducing the overall properties of various kinds of silicates and aluminosilicates \cite{Catlow1992Book}, and they have been applied for studying diffusion barriers of sodium vacancies in Na- and K-feldspar \cite{Jones2004}. These classical force fields, however, fall behind density functional theory (DFT) methods, which have been demonstrated to predict various properties of feldspar accurately \cite{Kaercher2014}. Yet, DFT and other {\em ab initio} methods are computationally considerably more expensive than classical force fields, drastically limiting the accessible system sizes and simulation times. Recently, machine learned force fields (MLFF) have been shown to offer accuracy comparable to first-principle electronic-structure calculations at a cost comparable to that of classical force-field calculations \cite{Behler2007, Csanyi2010}. Such MLFFs have already been used to study a vast array of systems ranging from organic molecules and inorganic crystals to surfaces, aqueous systems and biomolecules \cite{Unke2021,Friederich_2021,omranpour2024perspective}.
In particular, MLFFs have been used for studying diffusion in lead and cadmium telluride \cite{Minkowski2021} and an MLFF for studying ice nucleation on K-feldspar has been developed recently \cite{piaggi2023firstprinciples}.

In this paper, we develop a Behler-Parrinello type neural network potential (NNP) for sodium feldspar, paying particular attention to the accurate representation of defects. Trained on a reference data set obtained with DFT using the PBE functional \cite{Perdew1996PBE}, our NNP accurately reproduces forces and energies and it can be used to carry out extensive molecular dynamics simulations. We demonstrate its applicability by computing the heat capacity as well the elastic constants and the geometry of the unit cell of Na-feldspar over a wide range of temperatures, finding good agreement with experimental data. In addition, we investigate the structure and diffusion of charged interstitial and vacancy defects, taking electrostatic finite size corrections into account. Such Frenkel pairs, consisting of \ce{Na^+} interstitials and \ce{Na^+} vacancies, are expected to be the majority carrier of alkali defects, since Schottky defects are energetically improbable due to the high Si-O bonding energy~\cite{Behrens1990}. The computational efficiency of the NNP allowed us to determine the defect formation free energy, and from it the defect concentrations for temperatures from \SI{0}{\kelvin} up to \SI{1400}{\kelvin} using thermodynamic integration. Furthermore, our simulations revealed a new interstitial configuration, in which two \ce{Na^+} cations share an alkali lattice site. This dumbbell-like defect is thermodynamically more favorable than the \ch{Na^+_{\((00\tfrac{1}{2})\)}} interstitial that was considered previously \cite{Petrovic1974}. Interestingly, the energetically favourable orientation of the dumbbell changes discontinuously at a temperature of about \SI{752}{\kelvin}. The current version of our NNP can be used to model sodium feldspar, but it can be readily extended to potassium feldspar and the solid solution of alkali feldspar by retraining it with an extended traning set.

The remainder of the paper is organized as follows. In Sec.~\ref{sec:methods} we describe the computational methods used in our work, including the preparation of the reference data and the training of the NNP. Results are presented and discussed in Sec.~\ref{sec:results} and conclusions are provided in Sec.~\ref{sec:conclusions}.

\section{Methods}
\label{sec:methods}
\subsection{Ab-Initio Calculations}

All our DFT calculations were performed using the Vienna Ab-initio Simulation Package (VASP) \cite{Kresse1993VASP1, Kresse1996Vasp2, Kresse1996Vasp3} with the projector augmented wave method \cite{Kresse1999PAW} utilizing the Perdew-Burke-Ernzerhof (PBE) exchange-correlation functional \cite{Perdew1996PBE}. The conventional cell containing four formula units of \ce{NaAlSi3O8}, initially constructed using experimental data of Armbruster and colleagues \cite{Armbruster1990} with Al always kept at the T\(_1\)O-sites, was sampled using a \(6\!\times\!4\!\times\!6\) $\Gamma$-centered $k$-point mesh. The same $k$-point density in the Brillouin zone was kept across all calculations and the energy cutoff was set to \SI{560}{\eV}. The convergence criteria were \(10^{-5}\)\,\si{\electronvolt} for the self-consistent cycle and \(10^{-3}\)\,\si{\electronvolt\per\angstrom} for ionic relaxations. When varying the number of electrons, a neutralizing jellium background was included.

\subsection{Defect Formation Energies and their Correction}
\label{sec:DefectFormation}

To determine the equilibrium concentrations of point defects in Na-feldspar it is necessary to calculate defect formation energies. Since vacancies and interstitials carry a charge, their energy in a periodicically replicated system suffers from strong finite size effects that require proper corrections. For charged defects in a periodic supercell the defect formation energy is given by \cite{EFormationZhang1991,Freysoldt2014a}
\begin{equation}\label{Eq:DefectFormationEnergy}
\begin{split}
    E_{\text{form}} = &\left(E[\text{Defect}] + E_{\text{corr}}[\text{Defect}]\right) - E[\text{Bulk}]\\ 
    & - \sum_i n_i \mu_i + q (\epsilon_{\text{VBM}} +  \Delta E_{\text{FL}}).
\end{split}
\end{equation}
Here, \(E[\text{Defect}]\) is the total energy obtained from a supercell calculation containing the defect and \(E[\text{Bulk}]\) is the total energy of the pristine supercell. The term \(E_{\text{corr}}[\text{Defect]}\) is a finite-size correction explained in more detailed below. The integer \(n_i\) is the number of atoms of type \(i\) added to $(n_i>0)$ or removed from $(n_i<0)$ the system and \(\mu_i\) is the corresponding chemical potential in the reservoir with which the atoms are exchanged. The last term in the above equation represents the energy required to introduce or remove the electronic charge $q$, which is $q=-1$ for \ce{Na^+} vacancies and $q=+1$ for \ce{Na^+} interstitials. Here, \(\epsilon_{\text{VBM}}\) is the electron energy at the Valence Band Maximum (VBM) and \(\Delta E_{\text{FL}}\) the Fermi level. As is common for \SI{0}{\kelvin} DFT calculations, we identify \(\Delta E_{\text{FL}} = 0\).

As mentioned above, when calculating the formation energy of charged defects in the supercell approach one invariably introduces interactions between the defect, its periodic images and the neutralizing background charge, necessitating the correction $E_{\text{corr}}$. The most important contributions to this correction is the spurious Coulomb interaction, which remains significant for any practical supercell size. Several methods exist to subtract this interaction from the defective supercells. Apart from the extrapolative scheme, in which the defect formation energy is determined for a sequence of supercells of increasing size whose limiting value can be identified with the dilute limit, there have been major developments in formally removing the electrostatic interactions by corrections \cite{Kumagai2014}. Here, we use the correction of Kumagai and Oba \cite{Kumagai2014}. To compute this correction, we first determined the dielectric tensor of albite using DFT perturbation routines in VASP~\cite{Gajdos2006_dielectric_tensor}, see Tab.~\ref{Tab:dielectric_tensor}. Then, we calculated the correction from pristine and relaxed defective supercells of the various systems (listed in Tab~\ref{Tab:defect_form}) using the Spinney package \cite{Arrigoni2021spinney}. The potentials involved in the correction are provided in the Supplementary Information.

\begin{table}[!h]
    \centering
    \caption{Ionic (\(\epsilon^0\)) and electronic (\(\epsilon^\infty\)) contributions to the static dielectric tensor \(\epsilon_{ij}(0)\) of Na-feldspar computed in GGA. In order to compare with experimental data, we estimated the dielectric constant of a polycrystalline sample by \(\epsilon_{\textrm{poly}} = (\lambda_1 + \lambda_2 + \lambda_3)/3\), where \(\lambda_1, \lambda_2, \lambda_3\) are the eigenvalues of the dielectric tensor. The result is an \(\epsilon_{\textrm{poly}}\) equal to 6.18 versus values of 7 and 8 from experiments of Olhoeft~\cite{Olhoeft1989} and Jones et al.~\cite{Jones2004} respectively.}
    \begin{tabular}{c|S[table-format=2.4] S[table-format=1.4] S[table-format=1.4] }
                        & \(\epsilon^\infty\) & \(\epsilon^{0}\) & \(\Sigma\) \\\toprule
      \(\epsilon_{xx}\)  & 2.404 & 2.602 &   5.006  \\
      \(\epsilon_{yy}\)  & 2.441 & 4.664 &   7.105  \\
      \(\epsilon_{zz}\)  & 2.415 & 4.019 &   6.434  \\
      \(\epsilon_{yz}\)  & -0.005 & -1.576 & -1.581 \\
      \(\epsilon_{zx}\)  & -0.007 & -0.273 & -0.280 \\
      \(\epsilon_{xy}\)  & 0.001 & 0.110 &    0.111 \\ \hline
    \end{tabular}
    \label{Tab:dielectric_tensor}
\end{table}

\subsection{Architecture of the NNP-Committee}\label{sec:NNP_constr}

In this work, the potential energy surface is represented by artificial neural networks as proposed by Behler and Parrinello in 2007~\cite{Behler2007}. Local atomic environments are expressed by radial and angular atom-centered symmetry functions with a cutoff of \SI{6}{\angstrom}. The atomic feed forward neural networks consist of an input layer containing 256 nodes in the case of aluminum,  272 for silicon, 273 for sodium and 276 for oxygen. All four networks further consist of two hidden layers with 25 nodes each and one single output node. The weights and biases were initialized randomly and optimized using a parallel Kalman-Filter \cite{Singraber2019a} to minimize the root mean squared deviation between predicted energies and forces and the reference values. The n2p2 package \cite{Singraber2019} was used to train the potentials and to interface them with the molecular dynamics simulation package LAMMPS \cite{Thompson2022}.

We used several independent NNPs, trained with different initial weights, in a committee machine, {\em i.e.} predictions for a certain configuration \(A\) are calculated as an average over \(N\) independent predictions \(\mathbf{y}_i\):
\begin{equation}
    \overline{\mathbf{y}}(A) = \frac{1}{N} \sum_i \mathbf{y}_i (A).
\end{equation}
For all our calculations we used the committee method implemented in n2p2 by Kývala et al.~\cite{kyvala_committee} with a committee size of $N=4$. Committee machines are known to increase the precision of machine-learning models, reduce overfitting and they have been used to improve predictions of various physical properties \cite{NNPCommSchran2020,NNPCommImbalanzo2021_uncertaintyest}.
Furthermore, the variance \(\sigma(A)\) of the individual predictions can serve as a measure for the uncertainty of the averaged prediction \(\overline{\mathbf{y}}(A)\) \cite{NNCgenerrorWolpert1992, NNCgenerrorKrogh1994, NNCgenerrorUeda1996}. As explained below, we use this uncertainty in our active learning strategy for the generation of the reference data.

\subsection{Dataset Generation}
\subsubsection{Active Learning}
\label{sec:activelearning}

To generate the data set needed to train and test our NNP, we used a two-stage strategy. In the first stage, we carried out an on-the-fly machine learning simulation using VASP~\cite{OneTheFlyJinnouchi2019B, OnTheFlyJinnouchi2019}. In this approach, an {\em ab initio} molecular dynamics simulation is started and a Gaussian process regression (GPR) model is continuously trained on the energies and forces computed from DFT. As the simulation proceeds and the accuracy of the GPR model improves, the expensive electronic structure calculations are successively replaced by the computationally inexpensive GPR model. DFT calculations are only carried out if the accuracy of the GPR model, as determined from Bayesian error estimation, falls below a given threshold. Since with time the relevant part of configuration space gets explored by the simulation, fewer and fewer DFT calculations become necessary, speeding up the simulation considerably. 

With this on-the-fly approach, a relaxed \(2\!\times\!1\!\times\!2\) supercell was driven from 0 to \SI{1300}{\kelvin} in 20~000 steps at atmospheric pressure in the NPT ensemble. In these and all subsequent on-the-fly simulations a timestep of \SI{1.5}{\femto\second} was used. Whenever an electronic structure calculation was carried out, the respective configuration together with its energy and forces was added to the reference data set. The simulation was continued at \SI{1300}{\kelvin} until there were no new uncertain configurations added to the reference set for a span of at least 20~000 steps. The same procedure was repeated for a relaxed \(2\!\times\!1\!\times\!2\) supercell in which a sodium atom was introduced and an electron removed to create an \ch{Na^+_{i}} interstitial defect. This system took the longest to reach acceptable predictive accuracy at \SI{1300}{\kelvin} with 100~000 steps necessary. Finally, the procedure was carried out for a relaxed supercell in which a sodium was removed and an additional electron introduced to create a \ch{V^-_{Na}} vacancy. %
Throughout all of these simulations a total of 220~000 timesteps were carried out, long enough such that several intracrystalline hopping events of the defects were already observed therein. Of the total number of timesteps, 1679 were calculated using forces determined {\em ab initio}.

While MD simulations can be carried out with the GPR model directly, it is known that NNPs are superior to GPR models in terms of speed. In the second stage of our strategy, we have therefore used energies and forces computed {\em ab initio} for the 1679 configurations collected on-the-fly to train an NNP. More specifically, a committee of four neural-network potentials (their architecture is given in Sec.~\ref{sec:NNP_constr}) was trained with n2p2. The committee was then used to run MD simulations with \(2\!\times\!2\!\times\!2\) supercells in a pristine system and with interstitial or vacancy defects in the NPT ensemble up to \SI{1400}{\kelvin} with a timestep of \SI{1}{\femto\second}. Every 5th timestep the force uncertainty for each atom was determined. Since the force uncertainty is a local quantity in contrast to the uncertainty of the total energy, we expect it to be a more suitable measure of uncertainty. Configurations where any atom-wise force uncertainty was above the threshold of 5 times the RMSE (analogous to Ref.~\cite{NNPCommImbalanzo2021}) were collected. Configurations that were at least \SI{100}{\femto\second} apart were recalculated with VASP and put into the reference data set. With this augmented reference data set a new committee was trained and the process was repeated. After four such iterations the committee produced a low rate of uncertain configurations and the procedure was stopped. A total of 670 additional DFT calculations were necessary in this procedure. An increase in the required training set size between the initial GPR model of VASP and the final NNP had to be expected as GPR models are known to require less data than NNPs, see e.g. Ref. \cite{ZuoBehlerCsanyiPerformance2020}. In total, the training set contained 2351 configurations with energies and forces. 

To test the prediction of the NNP, a test set was created by sampling random structures throughout all simulations. In addition, with the final committee, MD simulations exceeding \SI{3}{\nano\second} of the \(2\!\times\!2\!\times\!2\) systems were performed and 400 structures each of the pristine system and the system containing an \ch{Na^+_{i}} as well as the system with a \ch{V^-_{Na}} were calculated with first principles and added to the test set. In total, the test set contained 1302 configurations with energies and forces. Let us denote the dataset obtained so far by
\begin{equation*}
    \mathcal{A}_1 = \{A_i, (E_i, \mathbf{f}_i) \},
\end{equation*}
where \(A_i\) are the configurations, and \(E_i\) and \(\mathbf{f}_i \) are the corresponding potential energies and forces, respectively, obtained in the supercell-calculations. 

\subsubsection{Including Corrections in the Dataset}

The dataset \(\mathcal{A}_1\) does not contain the finite-size corrections stipulated in Equation~(\ref{Eq:DefectFormationEnergy}) to calculate defect formation energies. 
Since machine-learning force-fields allow us to go to large system sizes at which the spurious long-range Coloumb interaction between a charged defect and its periodic images may actually become negligible, it was not clear if electrostatic finite-size effects should be included or corrected for in the reference data. To ensure a proper force-field that treats these charged defects consistently, we created a second labeled dataset:
\begin{equation*}
\begin{split}
    \mathcal{A}_2 &= \{A_i, (E_i + E_{\text{corr}}(A_i), \mathbf{f}_i) \},
\end{split}
\end{equation*}
where \(E_{\text{corr}}(A_i)\) is the electrostatic finite size correction that depends on the configuration \(A_i\). That is, corresponding to the system size and defect type, we add the correction before training to incorporate it into the neural-network potential. Note that the Kumagai-Oba correction is only defined for relaxed systems. We therefore identified \(E_{\text{corr}}(A_i)\) for any finite temperature configuration \(A_i\) by the correction of the relaxed system of the same supercell-size and defect type.

\subsection{Heat Capacity, Elastic Constants and Phonons}
\label{sec:methods_heat}

To assess the performance of the NNP compared to its DFT-reference as well as to experimental data we calculated the elastic constants and heat capacity. To calculate the elastic constants at the DFT-level we used the stress-strain relationship \cite{LePage2002} implemented in VASP. To calculate the elastic constants with NNPs we used the stress-strain relationship as implemented in LAMMPS/examples/ELASTIC. The isochoric heat capacity \(C_V\) was determined through the phonon density of states obtained using the phonopy \cite{togo2015Phonopy} codes for DFT and the NNPs respectively. To compare with experiments the isochoric heat capacity $C_V$ was transformed to the isobaric heat capacity $C_P$ through the relation \(C_P  = C_V  + T V \alpha^2 K\). Values for the expansion coefficient \(\alpha\) and the bulk modulus K were taken from literature \cite{Brown2006, Tribaudino2010}. A finite temperature anharmonic renormalization of the phonon band structure was performed using normal-mode decomposition \cite{Sun2014} as implemented in dynaphopy \cite{CARRERAS2017221}.

\subsection{Thermodynamic Integration and Parallel Tempering}

One of the goals of this work is to compute the equilibrium concentrations \( c(T)\) of interstitial and vacancy defects as a function of temperature. We accomplish this by determining the free energy of formation of Frenkel pairs \(G_\text{FP}(T)\), from which their concentration follows \cite{mehrer2007diffusion},
\begin{equation}\label{Eq:concentration}
    c(T) = \exp{\left(\frac{-G_\text{FP}(T)}{2k_\text{B}T}\right)}.
\end{equation}
The formation free energy \(G_\text{FP}(T)\) is calculated via thermodynamic integration  following Cheng and Ceriotti~\cite{Cheng2018} using i-PI~\cite{Kapil2019iPI} and LAMMPS~\cite{Thompson2022}. To achieve sufficient sampling also at low temperatures, we employ a parallel replica scheme \cite{Okabe2001} coupling isothermal-isobaric MD simulations at different temperatures, from 100K up to the melting point of \SI{1400}{\kelvin} \cite{Schairer1956, Dietz1970, Greenwood1998}. Note that this approach takes into account full anharmonic effects, which can become relevant at the higher temperatures \cite{VacAnhGrabowski2009,VacAnhGlensk2014}. Details about this free energy calculation are included in the Supporting Information.

\section{Results}
\label{sec:results}

\subsection{Performance of the Neural Network Potentials}

The accuracy of the neural network potential committees trained without charge corrections (NNPw/o) and with charge corrections (NNPw/corr) was first assessed by comparing the predicted forces and total energies with the reference data for the test set. The test set of NNPw/corr was appropriately shifted just like its training set \(\mathcal{A}_2\).

\begin{figure}[h]
    \centering
    \makebox[\columnwidth][c]{\includegraphics[width=0.83\columnwidth]{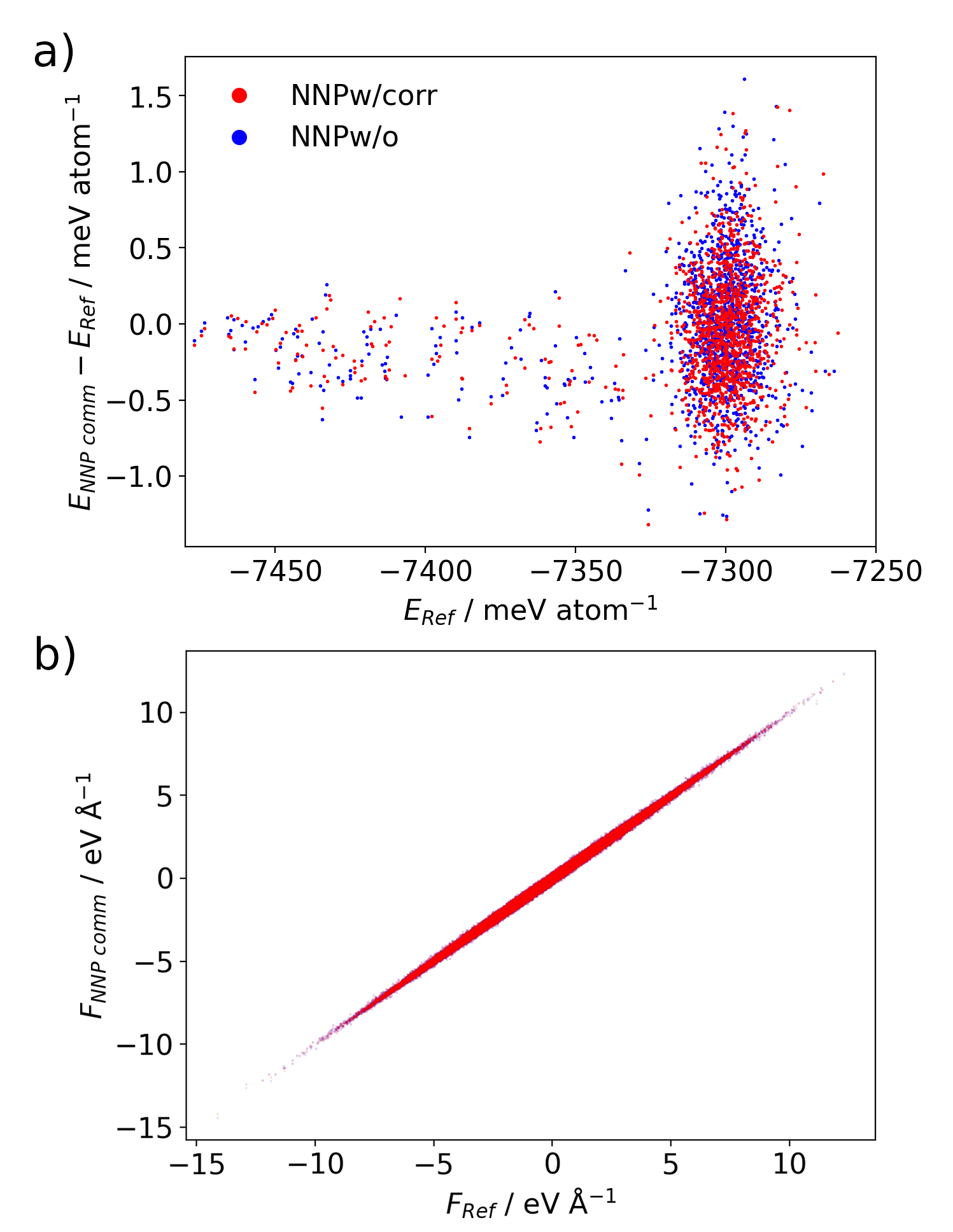}
    }
    \caption{ \(\mathbf{a})\) Deviation of the predicted energy from the reference energy vs. reference energy for NNPw/o and NNPw/corr, \(\mathbf{b})\) Predicted forces vs. reference forces for NNPw/o and NNPw/corr.
    }   \label{fig:NNP_Comp}
\end{figure}

As can be seen in Fig.~\ref{fig:NNP_Comp} and Tab.~\ref{Tab:rmse}, the NNP predictions are close to the reference data with an accuracy that is on par with similarly constructed neural-network potentials \cite{Minkowski2021} or other types of machine-learning force fields \cite{OnTheFlyJinnouchi2019}. In particular, the errors for NNPw/o and NNPw/corr are essentially the same. 

\begin{figure}[h]
    \centering
    \makebox[\columnwidth][c]{\includegraphics[width=1\columnwidth]{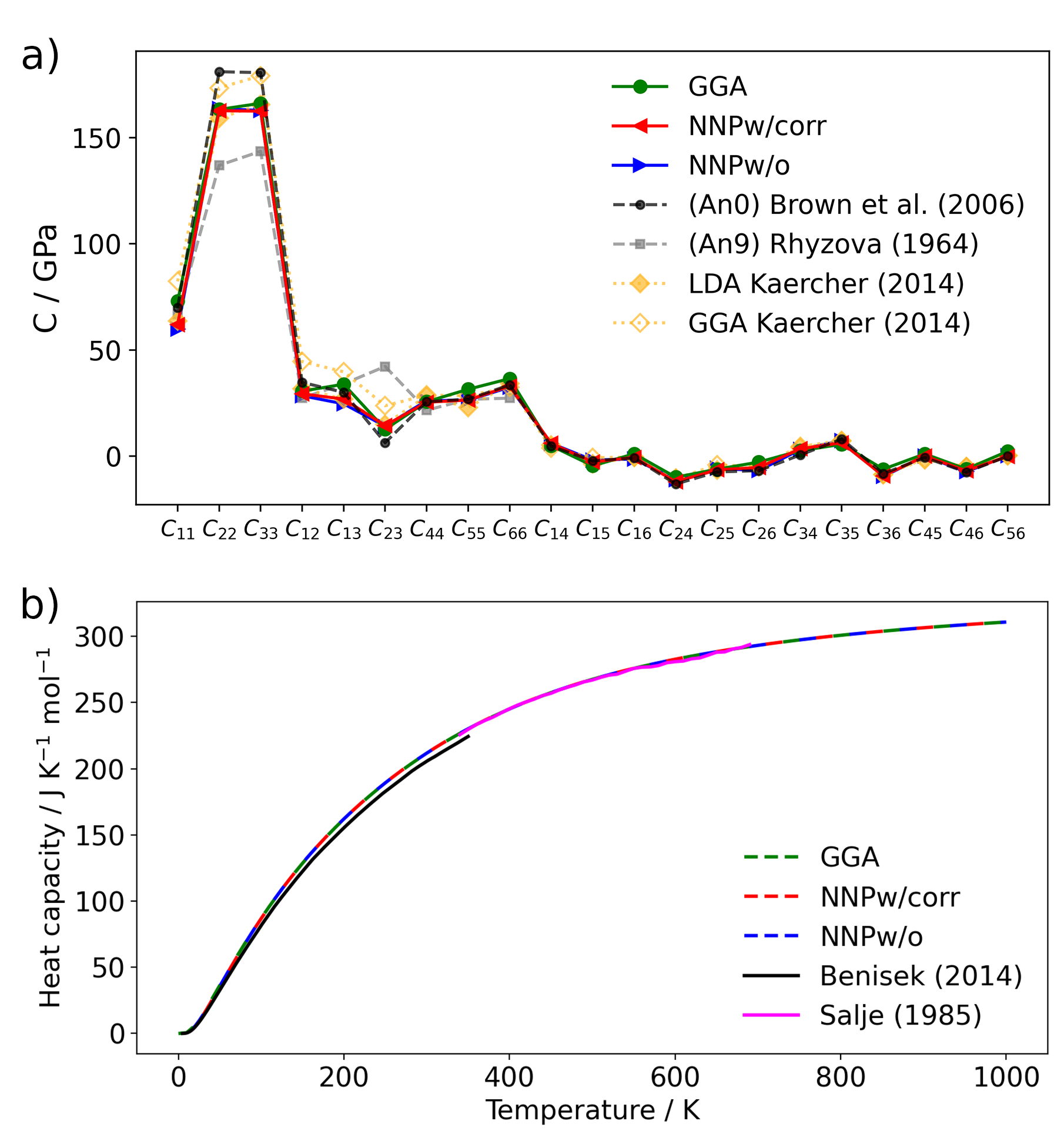}
    }
    \caption{ \(\mathbf{a})\) Elastic constants obtained from the NNP, DFT and experiments. Here we use the orientation convention \(\mathbf{Z}\|\mathbf{c}, \mathbf{Y}\|\mathbf{a}\times \mathbf{c}, \mathbf{X}\|\mathbf{Y} \times \mathbf{Z})\), ~\(\mathbf{b})\) Isobaric heat capacity $C_P$ vs. temperature.
    }   \label{fig:NNP_Properties}
\end{figure}

\begin{table}[!h]
    \centering
    \caption{Root mean squared errors for the test set obtained from the committee  trained without charge correction (NNPw/o) and and with charge corrections (NNPw/corr).}
    \begin{tabular}{l | c c }
    ~~~RMSE & NNPw/o & NNPw/corr \\\toprule
    \(\Delta E\) \(\left[\tfrac{\text{meV}}{\text{atom}}\right]\)   & 0.40   & 0.39 \\
    \(\Delta F\) \(\left[\tfrac{\text{eV}}{\si{\angstrom}}\right]\) & 0.065 & 0.065 \\
    \end{tabular}
    \label{Tab:rmse}
\end{table}

\subsection{Heat Capacity, Elastic Constants and Unit Cell}
In Fig.~\ref{fig:NNP_Properties}~(a) the elastic constants obtained from DFT as well as with the NNP with (NNPw/corr) and without (NNPw/o) corrections are shown. The figure also includes results of DFT calculations of Kaercher et al.~\cite{Kaercher2014} and experimental results of Brown et al.~\cite{Brown2006} and Rhyzova \cite{Ryzhova1964}. Note that in the calculations of  Kaercher et al.~\cite{Kaercher2014} the volume of the system was fixed at the experimental density during relaxation, making their results correspond to a high pressure. The results for NNPw/o and NNPw/corr almost fall on top of each other and both reproduce the DFT generated constants well overall. 

In Fig.~\ref{fig:NNP_Properties}~(b) the isobaric heat capacity obtained from DFT, NNPw/o, NNPw/corr and the experiments of Benisek et al.~\cite{Benisek2014} and Salje et al.~\cite{Salje1985} are shown. The heat capacities we obtained with three methods trace Salje's data, which are marginally higher than Benisek's experiments. 

\begin{figure}[h]
    \centering
    \makebox[\columnwidth][c]{\includegraphics[width=1\columnwidth]{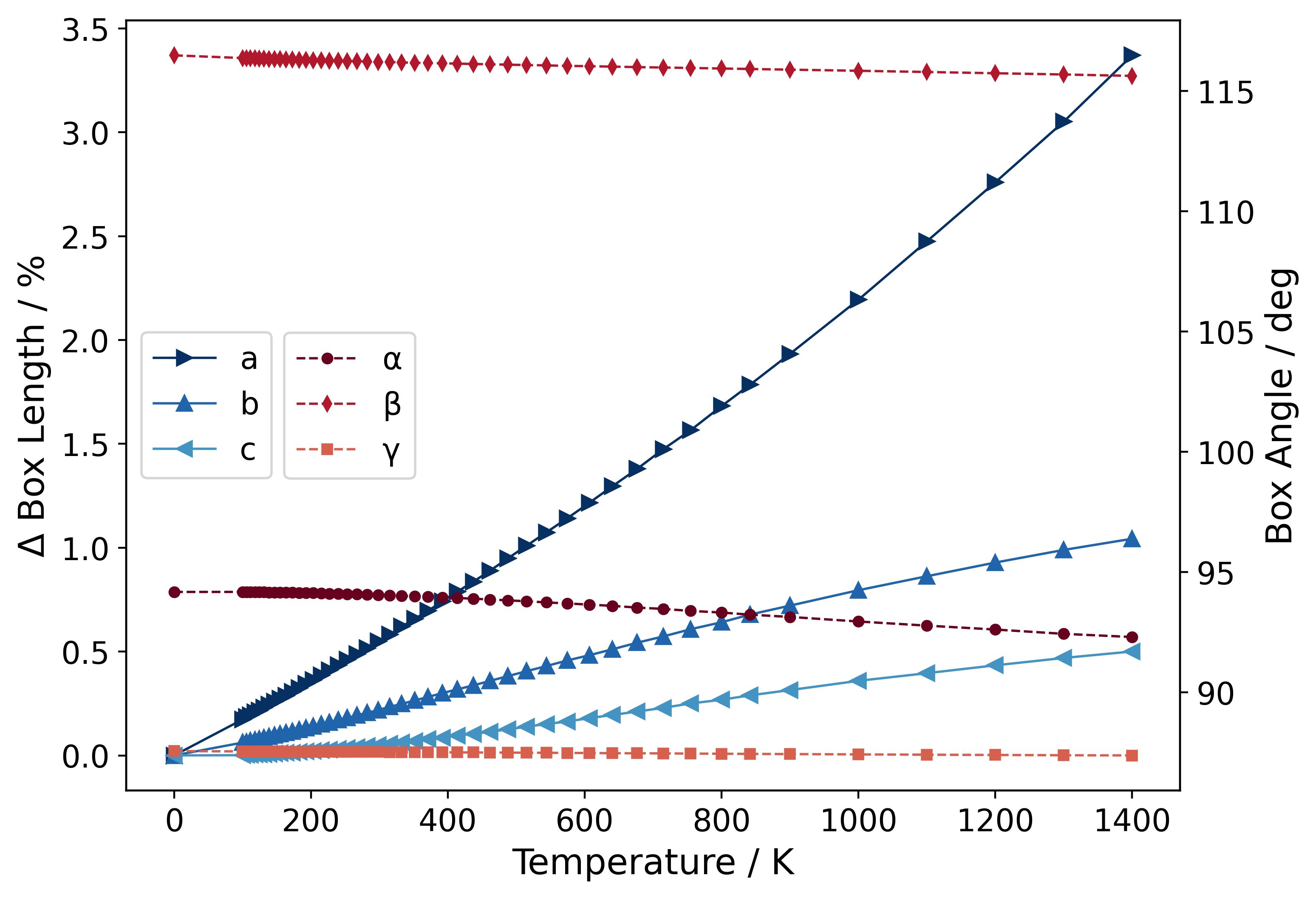
    }}
    \caption{Percent changes of the edge lengths $a, b, $ and $c$ (left $y$-axis) of the unit cell as a function of temperature with respect to the unit cell at 0K for NNPw/corr. Also shown are the angles $\alpha, \beta$ and $\gamma$ of the unit cell (right $y$-axis).
    }   \label{fig:cell_dim}
\end{figure}

In Tab.~\ref{Tab:unit_cell} we list the unit cell parameters obtained from  DFT, NNPw/o, and NNPw/corr as well as the experimental results of Brown et al.~\cite{Brown2006}. As is common with GGA-DFT, the lattice parameters are slightly overestimated, and this is reproduced in both NNPs. In Fig.~\ref{fig:cell_dim} we plotted the change in unit cell dimensions as a function of temperature for the NNPw/corr. As can be inferred from the figure, the lattice parameters increase anisotropically with temperature. The change of the box angles is in the correct direction as feldspars become more monoclinic as temperature increases \cite{Kroll1984}. %

\begin{table}[!h]
    \centering
    \caption{Unit cell parameters of Na-feldspar for the experiment of Brown and coworkers \cite{Brown2006} and fully relaxed unit cells of GGA and the two NNPs.}
    \begin{tabular}{l | c c c c c c}
        & a [\si{\angstrom}] & b [\si{\angstrom}] & c [\si{\angstrom}] & \(\alpha\) [\si{\deg}] & \(\beta\) [\si{\deg}] & \(\gamma\) [\si{\deg}] \\\toprule
        GGA &  8.260  &  12.933  &  7.250  &  94.210  &  116.489  &  87.555   \\
        NNPw/corr &  8.254  &  12.929  &  7.250  &  94.171  &  116.474  &  87.556   \\
        NNPw/o &  8.248  &  12.935  &  7.250  &  94.165  &  116.471  &  87.573  \\
        Brown et al.&  8.137   & 12.786   & 7.158 & 94.253  & 116.605  & 87.756   \\
    \end{tabular}
    \label{Tab:unit_cell}
\end{table}

\subsection{Defect Formation Energies at 0K}

\begin{figure}[b!]
    \centering
    \makebox[\columnwidth][c]{\includegraphics[width=1\columnwidth]{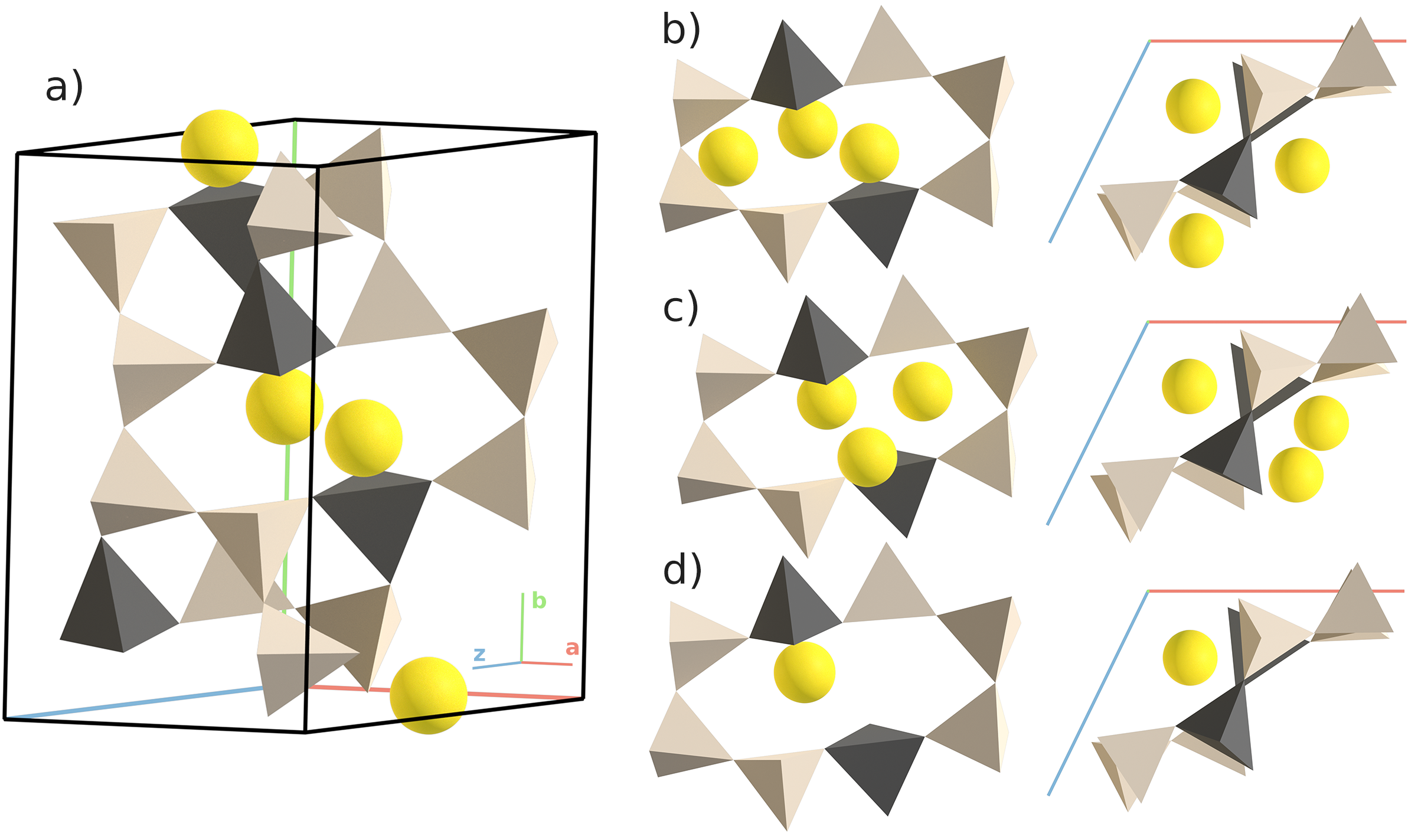
    }}
    \caption{Unit cell of albite (a) and three possible defects (b-d). White tetrahedrons are~\ce{[SiO4]^{-4}}, black ones are~\ce{[AlO4]^{-5}} and yellow spheres are~\ce{Na^+}. The defects are shown relative to the central eight-membered ring of tetrahedrons. The three defects, shown in side- and top-view, are: (b) the \ch{Na^+_{\((00\tfrac{1}{2})\)}} defect, (c) the dumbbell \ch{Na^+_{DB}} and (d) the vacancy \ch{V^-_{Na}}. 
    The configurations shown above were relaxed using NNPw/corr. Note that the unit cell used here is shifted by \((\tfrac{1}{2}\tfrac{1}{2}\tfrac{1}{2})\) with respect to the origin of the unit cell usually used for feldspar.
    }   \label{fig:defects}
\end{figure}

In the feldspar structure illustrated in Fig.~\ref{fig:defects}~a) an interstitial atom can occupy only a restricted number of positions. We started by placing a sodium cation at the \((0,0,\tfrac{1}{2})\) position, as proposed by Petrovic \cite{Petrovic1974}, and relaxed the system. This interstitial is refereed to as \ch{Na^+_{\((00\tfrac{1}{2})\)}} and is shown in Fig.~\ref{fig:defects}~b). In this configuration, the inserted cation lies on the same (010) plane as its neighboring sodium ions, to which it has a large separation. The two nearest sodium neighbors are \SI{4.17}{\angstrom} away and the two next-nearest are at a distance of \SI{4.34}{\angstrom}, such that the defect is at an electrostatically favorable position as argued by Petrovic. While this defect is stable at low temperatures, on heating the system as decribed in Sec. \ref{sec:activelearning} we observed that at a temperature of roughly \SI{600}{\kelvin} the defect transitions to the dumbbell-type defect \ch{Na^+_{DB}} in which two \ce{Na^+} share a single alkali lattice site, as shown in Fig.~\ref{fig:defects}~c). Such a dumbbell configuration is rare for ionic crystals due to electrostatic repulsion, but the large cavities of the feldspar framework seem to allow for this extra ion. The two \ce{Na^+} ions lie outside the (010) plane spanned by the other sodiums. They are only \SI{2.67}{\angstrom} apart but have a comparable separation to their next-nearest neighbors as conventionally occupied Na-sites. We also initialized a supercell with a sodium interstitial between two neighboring Na-sites inside the folded eight-membered ring as was suggested by Behrens and coworkers \cite{Behrens1990}. However, this configuration was unstable and reconfigured to the dumbbell configuration during relaxation. The final defect we discuss is the sodium vacancy \ch{V^-_{Na}} pictured in Fig.~\ref{fig:defects}~d).

The conventional unit cell of Na-feldspar contains 4 formula units in a C\(\overline{1}\) space group. A single formula unit supports only half of one \((0,0,\tfrac{1}{2})\)-site whereas the dumbbell and vacancy can appear once for every sodium site. The conventional unit cell therefore supports 2 \((0,0,\tfrac{1}{2})\)-sites, 4 vacancy-sites and 4 dumbbell-sites, each of them equivalent up to symmetry, respectively. We convinced ourselves of this explicitly for the dumbbell configuration and noticed that its axis remains the same on all possible locations. This implies that the two \ce{Na^+} are equidistant to the centre of inversion, which is situated in the middle of the eight-membered ring of tetrahedrons.

In Tab.~\ref{Tab:defect_form} we show the formation energies calculated for the defects in individual supercells as well as in systems containing both interstitial and vacancy defects together. The defect energies, calculated according to Equ. (\ref{Eq:DefectFormationEnergy}), were obtained for GGA, GGA including electrostatic corrections, the NNP committee trained on a reference without corrections (NNPw/o) as well as the NNP committee trained on a reference including corrections (NNPw/corr). 
The KO finite size charge correction for the charged defects is relevant in the \(2\!\times\!1\!\times\!2\) and only marginally less so in the \(2\!\times\!2\!\times\!2\) system. The corrected formation energy does however seem to be converged rather well already in the \(2\!\times\!2\!\times\!2\) system, decreasing only by \SI{0.03}{\electronvolt} compared to the smaller system for both \ch{V^-_{Na}} and \ch{Na^+_{\((00\tfrac{1}{2})\)}}.

To calculate the defect energy according to Equ. (\ref{Eq:DefectFormationEnergy}) we use the chemical potential of sodium bcc computed at \(T = 0\) and \(P = 0\) with GGA. The electron energy at the valence band maximum, $\epsilon_{\text{VBM}}$, of the perfect feldspar crystal was also determined with GGA under the same conditions. Note that in Ref.~\cite{Lowitzer2008}, GGA formation energies of sodium cation vacancies were computed without separating elemental and electronic contributions to the formation energy. Instead, the authors used the energy of a charged sodium ion in vacuum as reference, resulting in a defect formation energy that is \SI{9}{\electronvolt} higher than the corresponding energy shown in Tab. ~\ref{Tab:defect_form}.

The necessity of including NNPw/o and the NNPw/corr in this study becomes apparent when looking at the defect formation energies shown in the rightmost columns of Tab~\ref{Tab:defect_form}. Generally, NNPw/o and NNPw/corr seem to reproduce the uncorrected and corrected DFT reference, respectively. The precision at which this is accomplished, however, favors NNPw/corr. In particular, for the formation energy of the Frenkel pair, shown in the last row of Tab.~\ref{Tab:defect_form}, NNPw/corr predicts a value quite close to the corrected reference, while NNPw/o does neither predict the uncorrected energy nor the corrected one with acceptable precision. Note that this formation energy of the Frenkel pair is consistent with the upper limit of \(\sim \SI{2.4}{\electronvolt}\) of the enthalpic contribution to the formation free energy, which was deduced by El Maanaoui and coworkers~\cite{ElMaanaoui2016} using ionic conductivity measurements in K-rich feldspar.

\begin{table}[!h]
    \centering
    \caption{Defect formation energies at \(0\si{\kelvin}\) for various defects after full ionic and volumetric relaxation. The defect energies were calculated according to Equ. (\ref{Eq:DefectFormationEnergy}) using the internal energy per atom of sodium bcc, \(\mu_{Na} = -1.31~\si{\electronvolt}\), and a valence band maximum of \(\epsilon_{\text{VBM}} = 1.18~\si{\electronvolt}\), both determined in GGA. The Fermi level was set at the VBM, \(\Delta E_{\text{FL}} = 0\). The defect energy in the last row is that of the lowest energy Frenkel pair in the $2\times 2 \times 2$ system resulting from summation of the energies in rows 1 and 3.
    }
    \begin{tabular}{l c|c| S[table-format=3.2] S[table-format=2.2]| S[table-format=3.2] S[table-format=2.2]}
      &  &  & \multicolumn{4}{c}{Formation Energies \(E_\mathrm{form}\) [\si{\electronvolt}]}  \\
     &  &  &\multicolumn{2}{c|}{GGA}  &  \multicolumn{2}{c}{NNP} \\
    Defect Type  & System & \(q\)  & ~w/o  & w/corr  & ~w/o & w/corr   \\\toprule
    \ch{V^-_{Na}} & \begin{small}\(2\!\!\times\!\!2\!\!\times\!\!2\)\end{small} &  \num{-1} & 5.11   &  5.29           &   4.91   &   5.20       \\
    \ch{V^-_{Na}} & \begin{small}\(2\!\!\times\!\!1\!\!\times\!\!2\)\end{small} & \num{-1} & 5.03   &  5.32           &   4.91         &   5.21       \\
    \ch{Na^+_{DB}} & \begin{small}\(2\!\!\times\!\!2\!\!\times\!\!2\)\end{small} & +1 &  -3.34  &  -3.16      &   -3.38   &         -3.13     \\
    \ch{Na^+_{\((00\tfrac{1}{2})\)}} & \begin{small}\(2\!\!\times\!\!2\!\!\times\!\!2\)\end{small} &+1 &  -3.00  &  -2.82      &   -3.10         &   -2.83    \\
    \ch{Na^+_{\((00\tfrac{1}{2})\)}} &  \begin{small}\(2\!\!\times\!\!1\!\!\times\!\!2\)\end{small} &+1 &  -3.06  &  -2.79      &   -3.08         &   -2.81      \\
    \ch{V^-_{Na}}+\ch{Na^+_{DB}} & \begin{small}\(2\!\!\times\!\!2\!\!\times\!\!2\)\end{small}  &+0 &  1.87 &   \text{-}                   &      1.49       &    2.03     \\
    \ch{V^-_{Na}}+\ch{Na^+_{\((00\tfrac{1}{2})\)}} \vspace{1pt}  & \begin{small}\(2\!\!\times\!\!2\!\!\times\!\!2\)\end{small} &+0 &   2.27  &      \text{-}          &   1.79         &    2.35   \\ \hline
    \multicolumn{2}{c|}{Lowest energy Frenkel pair}  &+0 & 1.77  &    2.13         &   1.53   &     2.07   \\ \hline
    \end{tabular}
    \label{Tab:defect_form}
\end{table}
\twocolumngrid

\begin{table}[!h]
    \centering
    \caption{Defect Formation Energies calculated with GGA without altering the number of electrons in the supercells.}
    \begin{tabular}{l c|c| c}
    Defect Type ~~  & System & \(q\)  & \(E_\mathrm{form}\) [\si{\electronvolt}] \\\toprule
    \ch{V^0_{Na}} & \(2\!\!\times\!\!2\!\!\times\!\!2\)&+0 &   5.15   \\
    \ch{Na^0_{DB}} &  \(2\!\!\times\!\!2\!\!\times\!\!2\) &+0 &   1.53   \\
    \ch{Na^0_{\((00\tfrac{1}{2})\)}} & \(2\!\!\times\!\!2\!\!\times\!\!2\) &+0 &   1.91   \\
    \ch{Na^0_{\((00\tfrac{1}{2})\)}} \vspace{1pt}  & \(2\!\!\times\!\!1\!\!\times\!\!2\) &+0 &   1.92  \\\hline
    \multicolumn{2}{c|}{Lowest energy Frenkel pair} & +0 & 6.68 \\ \hline
    \end{tabular}
    \label{Tab:defect_form_no_charge}
\end{table}

The need for corrections is most visible in the defect formation energy of the Frenkel pairs in a single supercell, \(\big(\ch{V^-_{Na}} + \ch{Na^+_{DB}}\big)\) and \(\big(\ch{V^-_{Na}} + \ch{Na^+_{\((00\tfrac{1}{2})\)}}\big)\). As the Frenkel defect pair has no total charge, no charge correction is needed in this case. However, due to the finite size of the cell there remains an attraction between the interstitial defect and the vacancy, which decreases the formation energy compared to the sum of formation energies of \ch{V^-_{Na}} and \ch{Na^+_{DB}} or \ch{Na^+_{\((00\tfrac{1}{2})\)}}. In both cases, the NNPw/corr formation energy is closer to the reference than the NNPw/o result. 
Since no electrostatic correction needs to be applied in calculating the single Frenkel pair, and no configuration of a Frenkel pair in a single system was included in the training data, the good agreement of the NNPw/corr corroborates the validity of the electrostatic correction and indicates that it is a necessary ingredient in the construction of the training set. 

To ensure that we correctly identified the lowest energy charge state of the Frenkel pair defect, we also calculated the defect formation energies without altering the numbers of electrons and listed these energies in Tab.~\ref{Tab:defect_form_no_charge}. The last row shows the sum of uncharged \ch{V^0_{Na}} and \ch{Na^0_{DB}}, which corresponds to a Frenkel pair without charge transfer between vacancy and interstitial. The formation energy of this Frenkel pair is much higher than its counterpart with charge transfer shown in the last row of Tab.~\ref{Tab:defect_form}. Since the latter is also similar in value to the direct Frenkel pair \(\ch{V^-_{Na}} + \ch{Na^+_{DB}}\) in the single supercell, we conclude that it is the appropriate lowest energy charge state.

The energies shown in Tabs.~\ref{Tab:defect_form_no_charge} and \ref{Tab:defect_form} hold information on the transitions between charge states. The difference in formation energy of the interstitial in its neutral and charged state is \SI{4.69}{\electronvolt} in the case of the dumbbell and \SI{4.73}{\electronvolt} in the case of \ch{Na^+_{\((00\tfrac{1}{2})\)}}. An optical transition \(\left(\ch{Na^0_{DB}} + h \ch{->} \ch{Na^+_{DB}} + \hbar \omega\right)\) with \(h\) being a hole in the valence band would yield a photon with an energy of \SI{4.65}{\electronvolt}, where slight dissimilarities in the local lattice relaxations of \ch{Na^0_{DB}} and \ch{Na^+_{DB}} are taken into account. Possibly, this transition is related to a prominent peak in a recent XEOL spectrum of Na-rich feldspar at 4.5 to 4.3 \si{\electronvolt} (for 300 to 100 \si{\kelvin}) that has not been categorized yet \cite{Poolton2022}. Moreover, Garcia and coworkers \cite{Garcia1999} have detected a peak at \SI{4.3}{\electronvolt} and speculated that it originates from \ch{Na^+} at interfaces. Finally, a peak is also visible at \SI{4.5}{\electronvolt} in the phosphorescence spectrum of Na-feldspar \cite{Baril2003A}. An argument against a connection between \ch{Na^0_{DB} -> Na^+_{DB}} and those peaks is that the latter study also reports a small peak at \SI{4.4}{\electronvolt} in the phosphorescence spectra of 2 out of 4 specimens of K-rich feldspar. All these systems, however, contain also some amount of Na (as do all natural occurring Alkali feldspars) and we suspect that in Alkali feldspars of any composition, the dumbbell defect should be the most favorable interstitial defect as both K-K dumbbells and Na-K dumbbells would be sterically unfavorable compared to \ch{Na^+_{DB}}.

\subsection{Finite Temperature Behaviour}
\subsubsection{Free Energy of Defect Formation}

\begin{figure}[b!]
    \centering
    \makebox[\columnwidth][c]{\includegraphics[width=1\columnwidth]{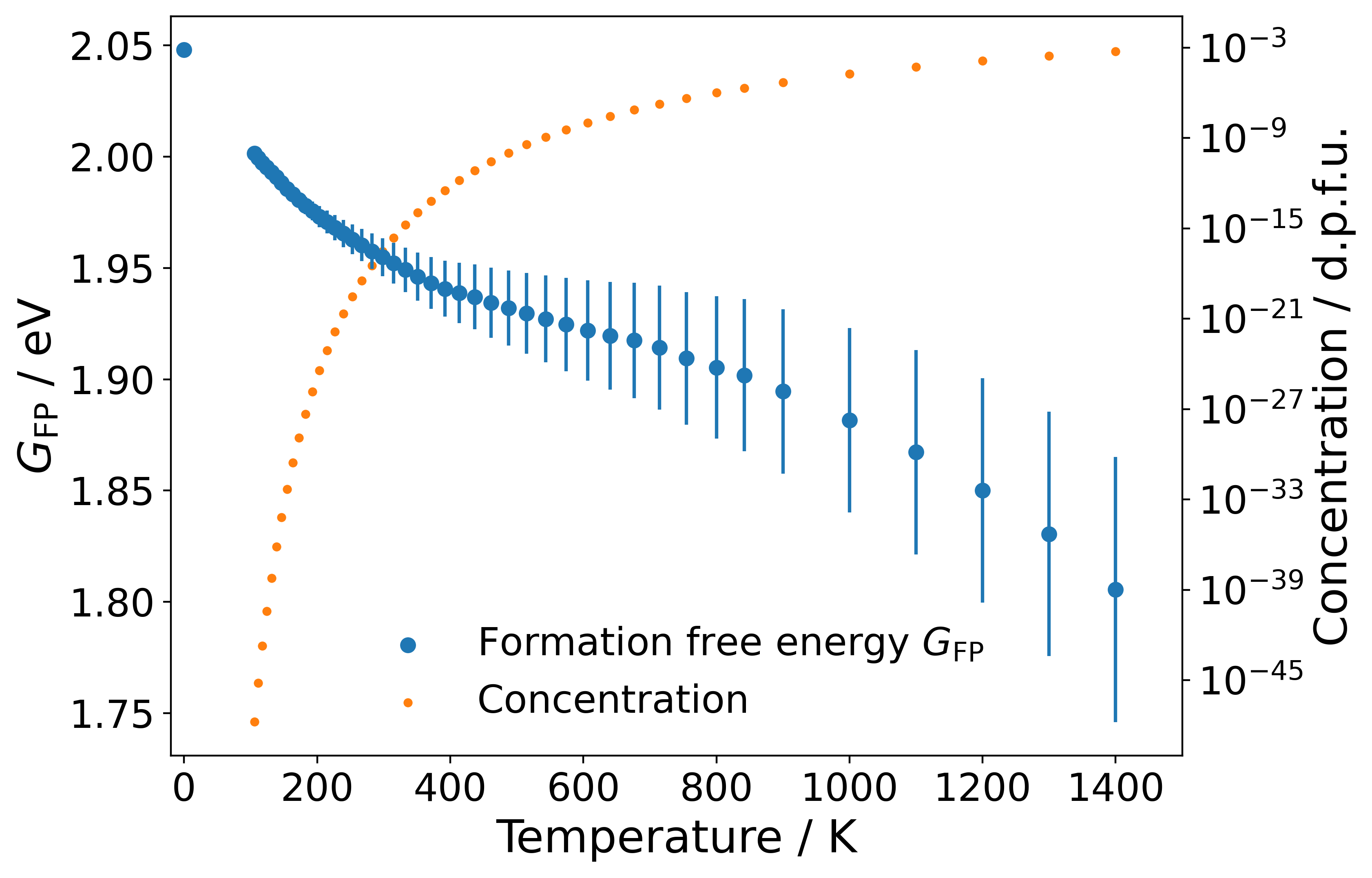}}
    \caption{Formation free energy $G_{\rm FP}$ with error bars of the Frenkel pair as a function o temperature $T$. The estimation of the error bars is described in the Supplementary Information. Also shown in the resulting defect concentration $c$ obtained by applying Eq.~\eqref{Eq:concentration}. Defect concentrations are given in defects per formula unit (d.p.f.u). 
    }   \label{Fig:free_energy_main}
\end{figure}

Using thermodynamic integration starting from a harmonic reference model we calculated the free energies at finite temperatures and atmospheric pressure for a pristine system, a system containing \ch{V^-_{Na}} and a system containing \ch{Na^+_{DB}}. Details of the thermodynamic integration procedure, combined with parallel tempering, and the resulting free energies are provided in the Supplementary Information. Based on these individual free energies we determined the free energy of formation of the Frenkel pair as a function of temperature
\begin{equation*}
    G_{\mathrm{FP}} = G_{\ch{V^-_{Na}}} + G_{\ch{Na^+_{DB}}} - 2~G_{\mathrm{pristine}},
\end{equation*}
which is plotted in Fig.~\ref{Fig:free_energy_main}. Interestingly, the free energy decreases monotonically with increasing temperature, but goes through an inflection point at around \SI{650}{\kelvin}. Below this temperature the free energy of formation decreases in a convex fashion and above it it decreases concavely. %

We note that at temperatures around and above the inflection point we observe transitions of the interstitial in its \ch{Na^+_{DB}} configuration to 
the \ch{Na^+_{\((00\tfrac{1}{2})\)}} defect. 
This defect state occurs very infrequently and when it does it is short-lived, quickly transitioning back to \ch{Na^+_{DB}}. Hence, the inflection point in the defect free energy might be related to additional configuration space volume that becomes accessible at higher temperatures.

\subsubsection{Discontinuity in the Dumbbell Orientation and a Dynamical Instability}

\begin{figure}[b!]
    \centering
    \makebox[\columnwidth][c]{\includegraphics[width=1\columnwidth]{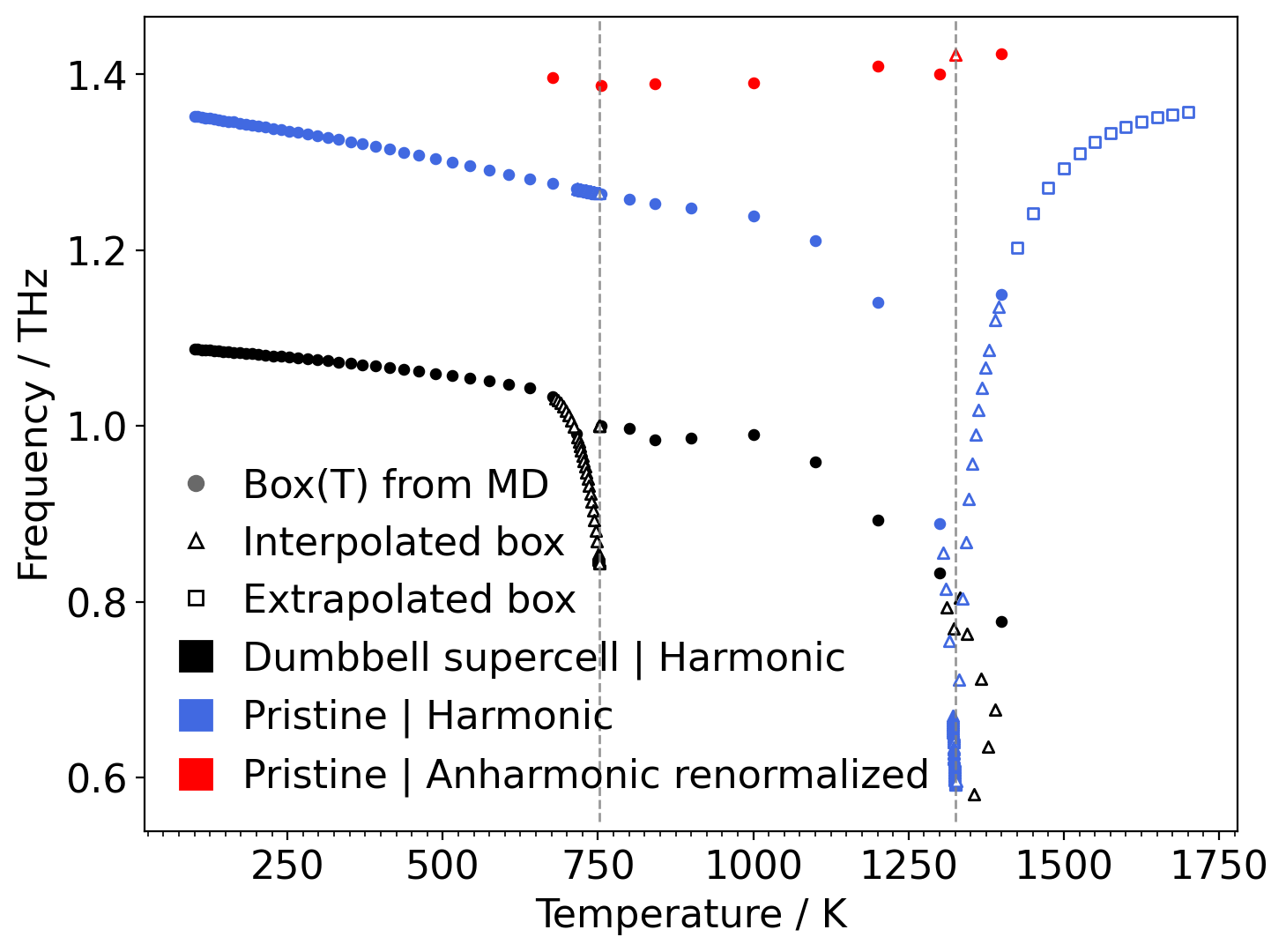}}
    \caption{The lowest normal mode frequency at the \(\Gamma\)-point. Here anharmonic renormalization refers to normal-mode-decomposition \cite{CARRERAS2017221}. The first minimum at \SI{751.91}{\kelvin} is related to a discontinuity in the orientational axis of the dumbbell. The soft mode at \SI{1325.60}{\kelvin} disappears when including anharmonic effects.
    }   \label{Fig:normalmodes}
\end{figure}

Further insights into the nature of the dumbbell defect can be obtained by analyzing the normal mode frequencies of the system with and without defect. 
In Fig.~\ref{Fig:normalmodes} we show the lowest normal mode frequency at the \(\Gamma\)-point of systems containing a dumbbell and of the pristine crystal as a function of temperature. We calculated the normal mode frequencies using the harmonic approximation or normal-mode decomposition, as described in Sec.~\ref{sec:methods_heat},
for the box dimensions corresponding to the respective temperatures (cf.~Fig.~\ref{fig:cell_dim}). The box dimensions as a function of temperature where determined from parallel tempering simulations. Close to the instability temperature discussed below additional box sizes were obtained by interpolation and at higher temperature by extrapolation.

In the system with the dumbbell, at a temperature of about \SI{752}{\kelvin} a frequency vanishes and reappears at a different value. Such a discontinuity implies that a minimum in the potential energy landscape disappears, after which the system relaxes to a different minimum. Since we do not observe the same discontinuity in the pristine system, we conclude that this effect is related to a mechanical instability of the dumbbell defect.
 
\begin{figure}[b!]
    \centering
    \makebox[\columnwidth][c]{\includegraphics[width=1\columnwidth]{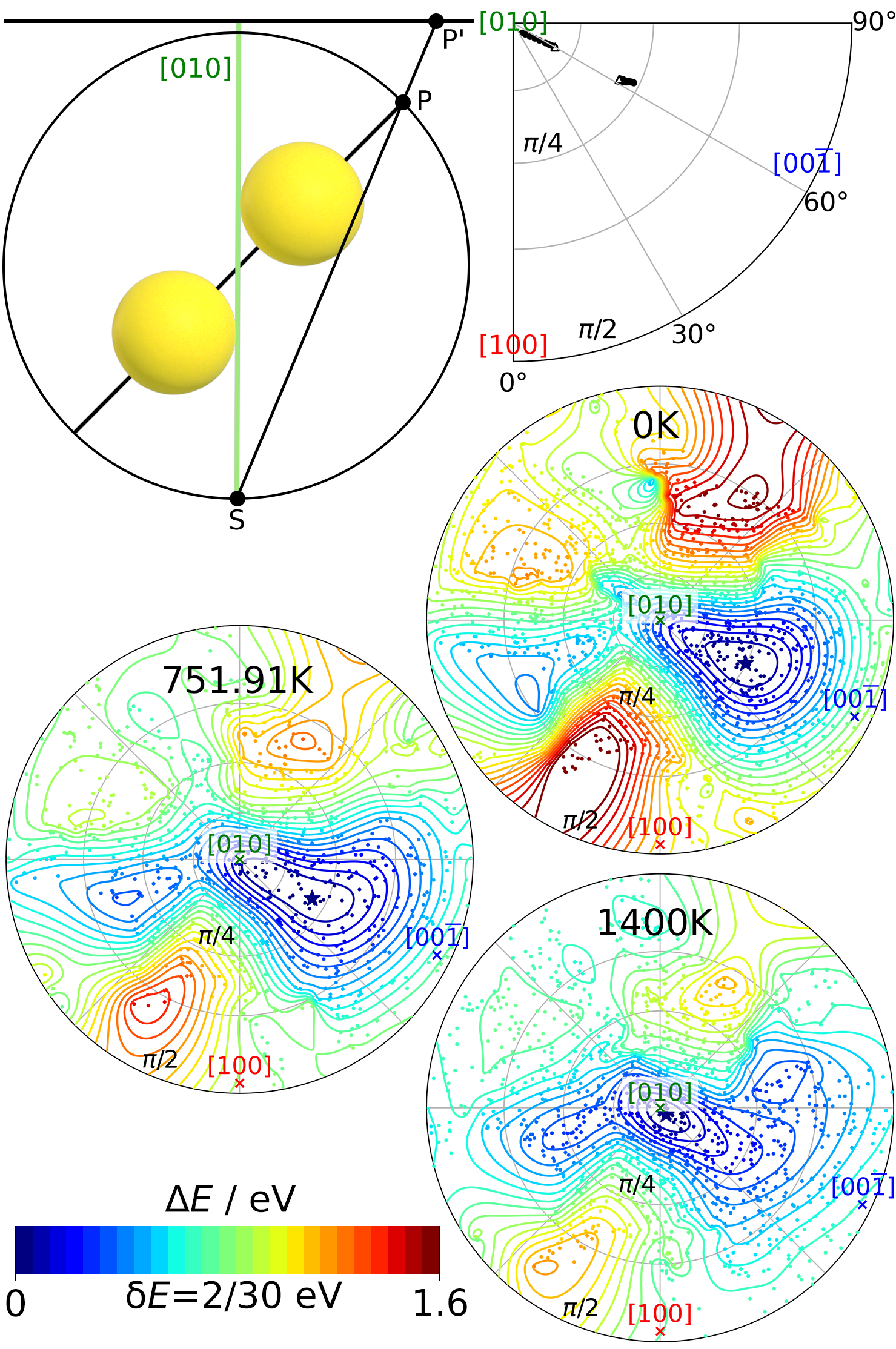}}
    \caption{The orientation of the dumbbell axis is shown using a stereographic projection, visualized in the upper left. To obtain this projection, the axis at \(P\) which lives on the upper hemisphere is mapped into the point \(P^\prime\) of the image plane. The \([010]\) direction is kept at \SI{0}{\radian} latitude and the \([100]\) at \SI{0}{\deg} longitude. In the upper right panel, the axis orientation corresponding to the energy minimum at all temperatures of Fig.~\ref{Fig:normalmodes} is plotted and the discontinuity in latitude occurring at \SI{751.91}{\kelvin} is evident. The lower part shows the potential energy landscapes sampled for different box dimensions. The axis orientation corresponding to the energy minimum is indicated by a star.
    }   \label{Fig:dumbbellaxis}
\end{figure}

A closer analysis reveals that the instability at \SI{752}{\kelvin} is related to the orientation of the axis of the dumbbell. Figure~\ref{Fig:dumbbellaxis}~(upper right) shows this axis at the potential energy minimum for all box dimensions in a stereographic projection. At \SI{0}{\kelvin} it has an almost \SI{45}{\deg} angle to the \([010]\)-direction (cf.~Fig.~\ref{fig:defects}~c). With increasing temperature this angle slightly decreases and at \SI{752}{\kelvin} it jumps by about \SI{22.5}{\deg}. Upon further temperature increase the angle decreases to almost \SI{0}{\deg} at \SI{1400}{\kelvin}. 

\begin{figure}[t!]
    \centering
    \makebox[\columnwidth][c]{\includegraphics[width=1\columnwidth]{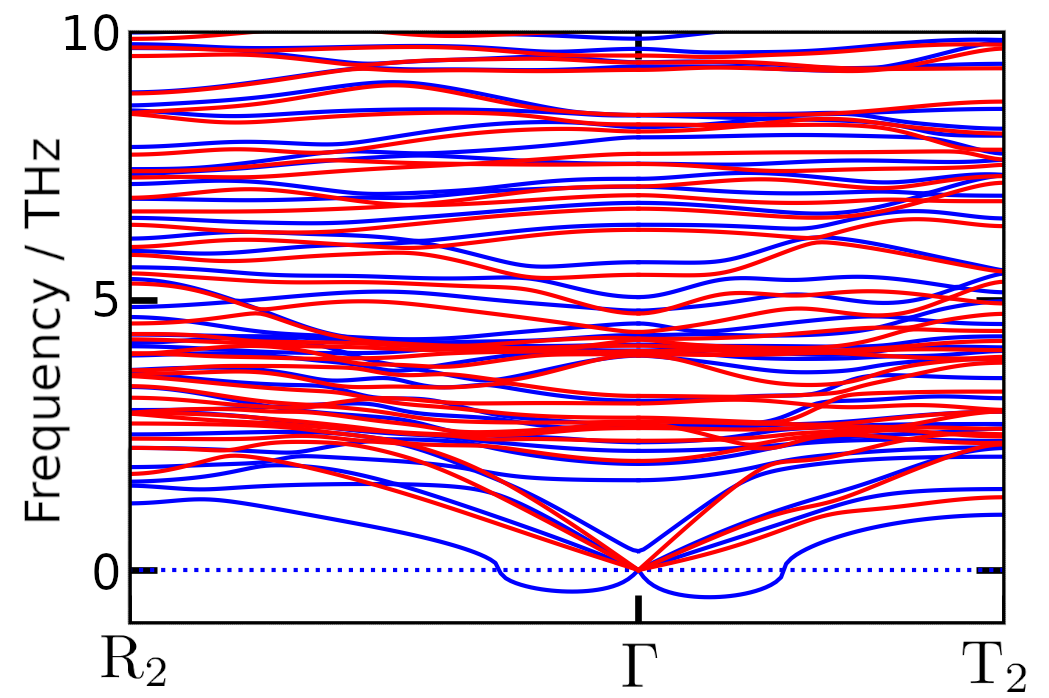}}
    \caption{Phonon bandstructure in the harmonic approximation (blue) and with anharmonicity included using normal-mode decomposition (red) at \SI{1325.60}{\kelvin}. Only a step of the band path suggested in \cite{HINUMA2017140} is shown. The complete band structure is included in the Supporting Information.
    }   \label{Fig:bandstructure}
\end{figure}

The lower parts of Fig.~\ref{Fig:dumbbellaxis} show the potential energy landscape as a function of the axis orientation for different temperatures. To determine these energy landscapes, we held the axis of the two \ce{Na^+} in the dumbbell fixed at random values and relaxed all other degrees of freedom (including the distance of the dumbbell atoms to each other). Only samples with a separation of the two \ce{Na^+} smaller than \SI{3.8}{\angstrom} are counted as dumbbell defects. Configurations with a larger separation correspond to a \ch{Na^+_{DB}} at a different alkali site or even an \ch{Na^+_{\((00\tfrac{1}{2})\)}} defect. 

Across all box dimensions the dumbbell axis shows high anisotropy and energetically favors orientations between \([010]\) and \([00\overline{1}]\). Orientations between \([010]\) and \([100]\) are energetically much more costly. As the box dimensions increase with temperature, the well between \( [010]\) and \([00\overline{1}]\) flattens and at \SI{752}{\kelvin} a new minimum emerges. As only one  alkali site is considered, the potential energy landscape would have to be appropriately inverted for an \ch{Na^+_{DB}} at the alternative alkali site.

The necessity for using thermodynamic integration to incorporate anharmonic effects in the free energies of Fig.~\ref{Fig:free_energy_main} is evidenced by the appearance of a soft mode at about \SI{1325}{\kelvin} in the harmonic approximation in both pristine and interstitial systems, as shown in Fig~\ref{Fig:normalmodes}. Below \SI{1300}{\kelvin} the crystal is dynamically stable, but between 1300-1400~\si{\kelvin} an optical mode almost vanishes and an acoustic band exhibits a dynamical instability. An excerpt of the bandstructure at \SI{1325.60}{\kelvin} is shown in Fig.~\ref{Fig:bandstructure}. Curiously, the crystal regains its harmonic stability if we extrapolate the box dimensions beyond \SI{1400}{\kelvin}. The dynamical instability in the bandstructure disappears completely after renormalizing the bands by incorporating finite temperature effects using normal-mode decomposition. This shows that the dynamical instability is a harmonic artifact and that anharmonic effects are essential in modelling Na-feldspar at these high temperatures.

\section{Conclusions}
\label{sec:conclusions}

In this work we presented an NNP, purpose-built for Na-feldspar (Albite) and its defects, and demonstrated its applicability to simulate this system by predicting relevant physical properties. In constructing the potential we made use of on-the-fly machine learning initially and then expanded the dataset using an NNP committee machine. This way to proceed required less human intervention than typically needed for NNPs and should be widely applicable.

We have shown that electrostatic corrections need to be included into the training dataset for the NNP. This is  particularly important, when defect formation energies are of interest. Including corrections after training produces a potential that cannot transfer to systems that are made up of multiple defects that neutralize each other, such as Frenkel pairs in Na-feldspar.

Based on our NNP, we have found a new kind of interstitial defect in the dumbbell configuration \ch{Na^+_{DB}}, where two \ce{Na^+} cations share a single site on the alkali sub-lattice. This defect is the energetically most favorable, but the \ch{Na^+_{\((00\tfrac{1}{2})\)}} interstitial predicted previously also appears at elevated temperatures. These different defect configurations as well as the instability in the axis orientation of the dumbbell defect might affect the free energy of formation of the Frenkel pair, which we found to go through an inflection at elevated temperatures.

A promising future application of the NNP develped in our work will be the direct determination of diffusion coefficients of defects using MD. Combined with the free energy of defect formation, it will be possible to compare these with experimentally determined tracer diffusion coefficients and elucidate the underlying diffusion mechanism. A second topic of interest is the extension of this NNP potential to include also potassium feldspar to study mixed systems of both sodium and potassium feldspar, if possible including disorder in the sites occupied by silicon and aluminum. These points will be addressed in future research. 

\section{Acknowledgments}
We would like to thank Ferenc Karsai and the VASP team for supporting us in using the on-the-fly methododology in its then development stage. We further acknowledge financial support from the Austrian Science Fund (FWF) through Grant No.~I 4404 and through the SFB TACO, Grant No.~F-81. The computational results presented were achieved using the Vienna Scientific Cluster (VSC).

\section{Data Availability}
The training and test data, the NNPw/corr, a template for molecular dynamics as well as LAMMPS-DATA configurations of the pristine, \ch{Na^+_{\((00\tfrac{1}{2})\)}}, \ch{Na^+_{DB}}, \ch{V^-_{Na}}, \ch{V^-_{Na}}+\ch{Na^+_{DB}} and \ch{V^-_{Na}}+\ch{Na^+_{\((00\tfrac{1}{2})\)}} systems are openly available through the Zenodo repository (\href{https://doi.org/10.5281/zenodo.10630971}{https://doi.org/10.5281/zenodo.10630971}).

\pagebreak
\end{document}


\onecolumngrid
\begin{center}
\textbf{\large Supplemental Materials: Structure and thermodynamics of defects in Na-feldspar from a neural network potential}
\end{center}
\setcounter{section}{0}
\renewcommand{\thesection}{S-\Roman{section}}

\section{Electrostatic Correction of Kumagai and Oba}
\begin{figure}[h!]
    \centering
    \makebox[\columnwidth][c]{\includegraphics[width=0.75\columnwidth]{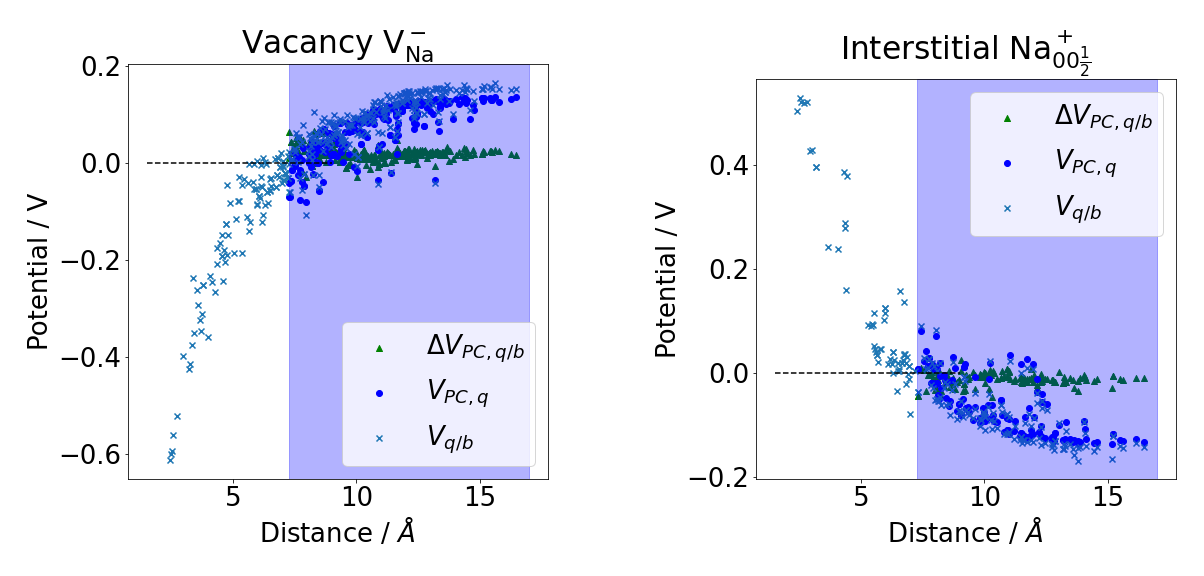}}
    \caption{Values of the different potentials involved to calculate the correction as functions of their distance to the defect. Colored in blue is the region used in sampling.}   \label{fig:ko_potentials}
\end{figure}

The potentials involved in the KO-correction, defined in \cite{Kumagai2014}, are plotted using the Spinney package \cite{Arrigoni2021spinney} in Fig.~\ref{fig:ko_potentials} for \ch{V^-_{Na}} and \ch{Na^+_{\((00\tfrac{1}{2})\)}} in the \begin{small}\(2\!\times\!2\!\times\!2\)\end{small} system. In both cases the difference between the defect-induced potential \(V_{q/b}\) to the point-charge potential \(V_{\mathrm{PC},q}\) reaches a constant value inside the sampling region. Both charges seem to possess a similar degree of localization as their respective defect-induced potentials mirror each other.
\section{Thermodynamic Integration}
\begin{figure}[h!]
    \centering
    \makebox[\columnwidth][c]{\includegraphics[width=0.5\columnwidth]{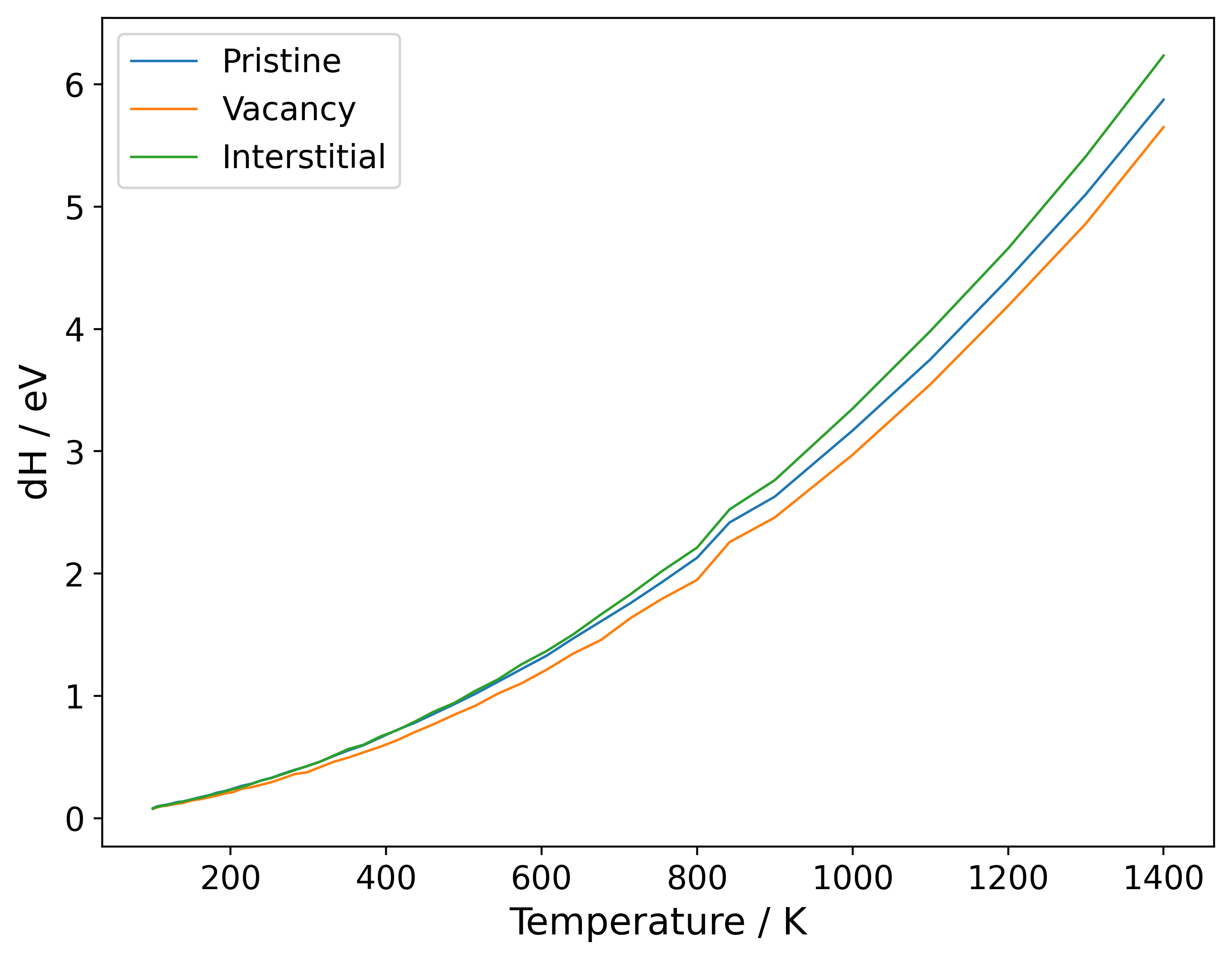}}
    \caption{Anharmonic contribution to the enthalpy.}   \label{fig:h_def}
\end{figure}

\begin{figure}[h!]
    \centering
    \makebox[\columnwidth][c]{\includegraphics[width=0.5\columnwidth]{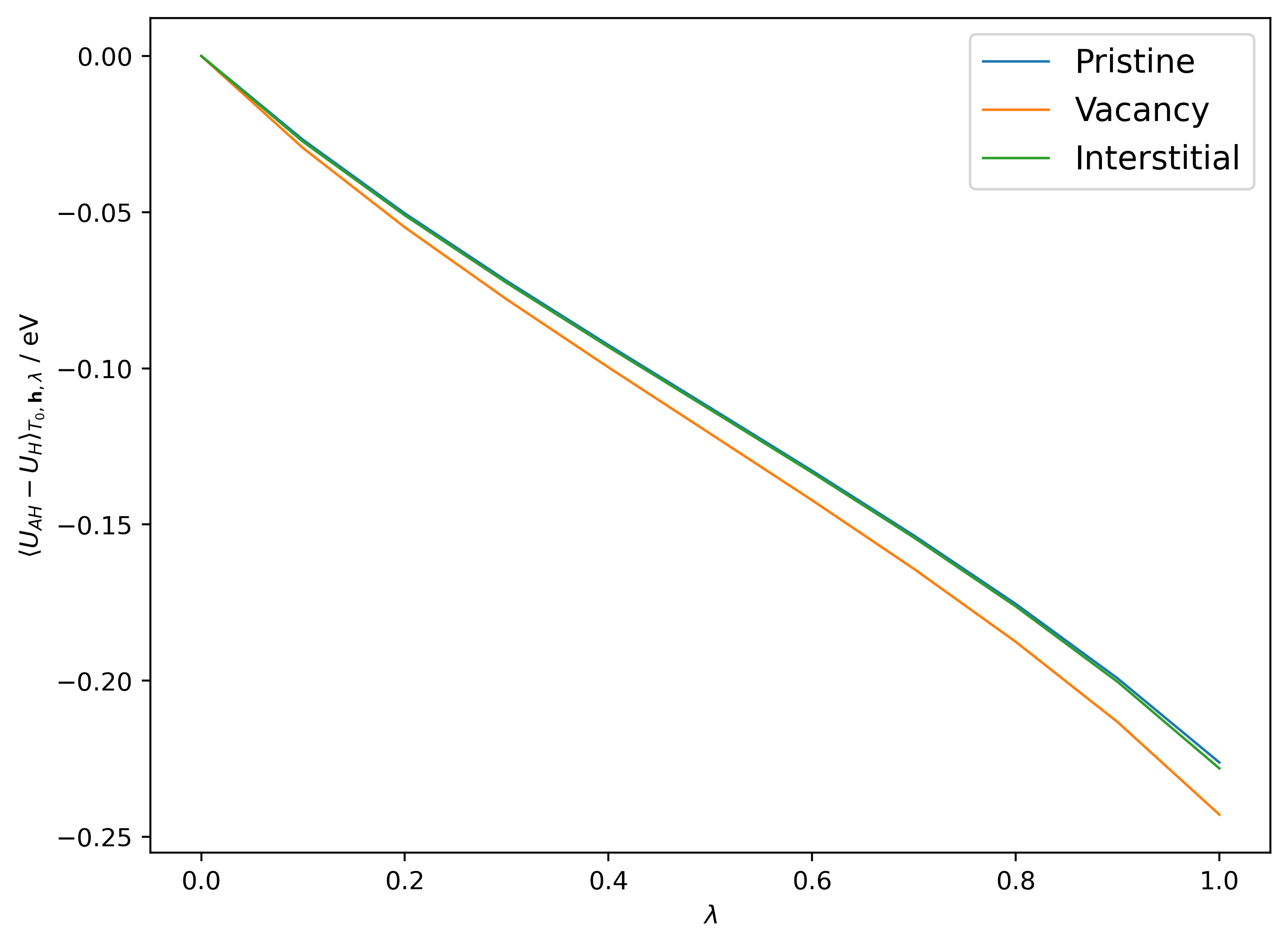}}
    \caption{Average difference between potential energies of harmonic and anharmonic hamiltonians in the ensemble of the effective Hamiltonian determined by \(\lambda\).} \label{fig:aanharm}
\end{figure}

\begin{figure}[h!]
    \centering
    \makebox[\columnwidth][c]{\includegraphics[width=0.5\columnwidth]{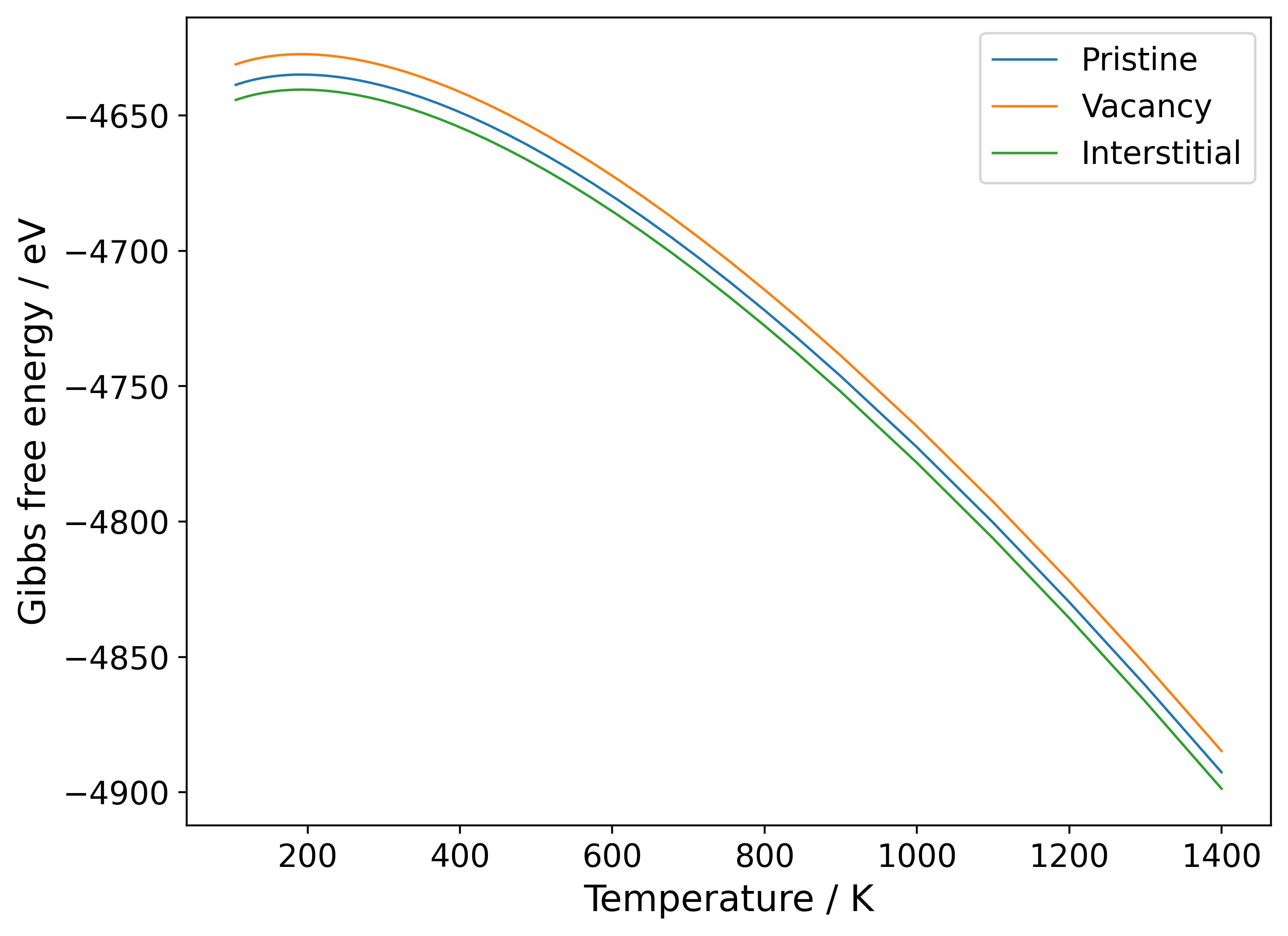}}
    \caption{Gibbs free energies.} \label{fig:gibbs_free_energies}
\end{figure}

Thermodynamic integration was performed using the approach described by Cheng and Ceriotti \cite{Cheng2018}, except that we have performed it at atmospheric pressure and have coupled the simulations at different temperatures with parallel tempering. Without parallel tempering, the free energy of defect formation at low temperatures did not converge properly. We provide a short account of the process below, which we repeated for a pristine system measuring \begin{small}\(2\!\times\!2\!\times\!3\)\end{small} unit cells, the same with a vacancy \ch{V^-_{Na}} and once more with the dumbbell interstitial \ch{Na^+_{DB}}. The real/anharmonic potential of interest is given by NNPw/corr.
\begin{enumerate}
    \item \textbf{Ionic and box degrees of freedom were relaxed at 0~K}. Relaxing the triclinic box required some care. We used a conjugated gradient descent until the energy varied less than \(10^{-10}\)~\si{\electronvolt} and continued with a steepest gradient descent until the energy varied less than \(10^{-12}\)~\si{\electronvolt}. The minimization was done with LAMMPS at atmospheric pressure, resulting in free energies \(G(P = \SI{1.01325}{\bar}, T = \SI{0}{\kelvin})\).
    \item \textbf{NPT simulations were performed between 100~K to 1400~K at atmospheric pressure with parallel tempering}. The replicas were equilibrated at the chosen temperatures in Tab.~\ref{Tab:free_energy_frenkel_pair} with parallel tempering in the NPT ensemble \cite{Okabe2001} for \SI{1}{\nano\second}. This corresponded to about twice the time the simulation that started at \SI{841.80}{\kelvin} initially, took to reach the lowest temperature through exchanges. The run continued for at least \SI{4}{\nano\second} more in which produced energies and boxes were saved every \SI{10}{\femto\second}. The acceptance rate was kept approximately equal from \SI{100}{\kelvin} to \SI{841.80}{\kelvin} by distributing these target temperatures exponentially. From \SI{900}{\kelvin} to \SI{1400}{\kelvin} they were distributed linearly instead, since simulations at these temperatures can be expected to be well-equilibrated by themselves already and to allow for a more dense distribution at the high temperatures which are of increased interest to study diffusion. Swaps were attempted every~\SI{0.1}{\pico\second}, the timestep was \SI{1}{\femto\second} and the center of mass was constrained in all replicas. The anharmonic enthalpy, defined in  \cite{Cheng2018}, was determined and plotted in Fig.~\ref{fig:h_def}.
    \item \textbf{Equilibrium configuration and phonon modes at 100~K} The ionic degrees of freedom of a system with the averaged box dimension \(\mathbf{h}\) for the replica at \SI{100}{\kelvin} were relaxed, and then the Hessian matrix at this minimum was determined with i-PI \cite{Kapil2019iPI}. The vibrational modes of the Hessian were then used to describe a reference harmonic crystal with a known free energy to allow determining the free energy of the anharmonic potential NNPw/corr by thermodynamic integration from the reference.
    \item \textbf{Integration from the reference harmonic to the real anharmonic potential in NVT.} We switched from the reference harmonic hamiltonian \(\mathcal{H}_h\) to the anharmonic NNPw/corr  \(\mathcal{H}_{a}\) at \SI{100}{\kelvin} by increasing \(\lambda\) from 0 to 1 in the effective Hamiltonian:
    \begin{equation*}
         \mathcal{H}(\lambda) = (1 - \lambda)\mathcal{H}_h + \lambda \mathcal{H}_{a}.
    \end{equation*}
    The integration was done in steps of \(\Delta\lambda = 0.1\) and the results are illustrated in Fig.~\ref{fig:aanharm}.
    \item \textbf{Putting everything together.} We used the annotated script provided in the SI of \cite{Cheng2018} to obtain the Helmholtz free energy at \SI{100}{\kelvin}. To transform it to the Gibbs free energy at the same temperature we had to extend the procedure in the script to triclinic systems. The Gibbs free energy is obtained from:
    \begin{equation*}
        G(P,T) = A(\mathbf{h},T) + T \mathrm{det}(\mathbf{h}) + k_B T ~\mathrm{ln}\rho(\mathbf{h}|P,T),
    \end{equation*}
    where \(\rho(\mathbf{h}|P,T)\) is the probability to observe the cell parameters \(\mathbf{h}\) in the ensemble of \(P\) and \(T\). To determine this probability from the results of recorded box sizes we first determined the deviations of recorded box dimension at the NPT replica at \SI{100}{\kelvin} to their average \(\mathbf{h}\) for each of the 6 degrees of freedom, resulting in different realizations of
    \begin{equation*}
        \Delta \mathbf{h} = \begin{bmatrix}
                                \Delta XX  & 0 &0 \\
                                \Delta XY & \Delta YY  & 0 \\
                                \Delta XZ & \Delta YZ & \Delta ZZ \\
                            \end{bmatrix}.
    \end{equation*}
    The probability was then determined by calculating the Frobenius norm \(||\Delta \mathbf{h}|| = \sqrt{\Sigma_{ij}\Delta \mathbf{h}_{ij}^2}\) for each realization and approximating \(\rho\) as:
    \begin{equation*}
         \rho(\mathbf{h}|P,T) \approx \sum^N_i \frac{\mathbbm{1}_{x < r }(\Vert \Delta \mathbf{h}_i \Vert)}{N \cdot V_{6~\text{ball}}(r)},
    \end{equation*}
    in which \(\mathbbm{1}_{x < r }(x)\) is the identity for values \(x\) smaller than \( r\) and zero otherwise. The expression is a histogram in which we count deviations whose Frobenius norm is smaller than \(r\) as realizations of \(\mathbf{h}\), normalized by all possible realizations. This is identical to the procedure taken in \cite{Cheng2018} for the orthogonal box, except with the addition of the three additional box degrees of freedom and the appropriate \(V_{6~\text{ball}}\) instead of \(V_{3~\text{ball}}\). The radius of the ball \(r\) in which we count the deviation as a realization was varied until a local maximum was found. Using the 2-norm instead of the Frobenius produced an almost identical result. The difference of the term \(
     k_B T ~\mathrm{ln}\rho(\mathbf{h}|P,T)\) between pristine, \ch{V^-_{Na}} and \ch{Na^+_{DB}} was less than \SI{2}{\milli\electronvolt} in any case. \\
    We could then carry on with the process \cite{Cheng2018} by including a correction of the constrained center of mass, integrating the free energies over temperature resulting in Fig.~\ref{fig:gibbs_free_energies}  and finally generating the free energies of formation for the Frenkel pair given in Tab.~\ref{Tab:free_energy_frenkel_pair} with the relation:
    \begin{equation*}
        G_{\mathrm{frenkel~pair}} = G_{\mathrm{vacancy}} + G_{\mathrm{interstitial}} - 2~G_{\mathrm{pristine}},
    \end{equation*}
    which does not contain temperature dependent chemical potentials, which would be needed to calculate the free energy of formation for the individual defects separately.\\
\end{enumerate}

\begin{table}[!h]
    \centering
    \caption{Gibbs free energies of formation of the frenkel pair at atmospheric pressure.}
    \begin{tabular}{c c c || c c c }
        \(T\) [K] & \(G_{\mathrm{FP}}\) [eV] & Uncertainty [eV]~ & \(T\) [K] & \(G_{\mathrm{FP}}\) [eV] & Uncertainty [eV] \\\toprule
        105.614  &  2.0015  &  0.0007  & 370.990  &  1.9432  &  0.0117  \\
        111.544  &  1.9995  &  0.001   & 391.819  &  1.9407  &  0.0126  \\
        117.807  &  1.9973  &  0.0013  & 413.817  &  1.9389  &  0.0136  \\
        124.421  &  1.9952  &  0.0016  & 437.051  &  1.937  &  0.0146  \\
        131.407  &  1.9931  &  0.0019  & 461.589  &  1.9344  &  0.0157  \\
        138.784  &  1.9909  &  0.0022  & 487.505  &  1.932  &  0.0169  \\
        146.576  &  1.9883  &  0.0025  & 514.875  &  1.9296  &  0.0182  \\
        154.806  &  1.9856  &  0.0028  & 543.783  &  1.9271  &  0.0195  \\
        163.497  &  1.9831  &  0.0031  & 574.313  &  1.9246  &  0.021  \\
        172.677  &  1.9806  &  0.0035  & 606.558  &  1.922  &  0.0225  \\
        182.372  &  1.9781  &  0.0039  & 640.613  &  1.9195  &  0.0242  \\
        192.611  &  1.9756  &  0.0043  & 676.580  &  1.9175  &  0.0259  \\
        203.425  &  1.9731  &  0.0047  & 714.566  &  1.9142  &  0.0278  \\
        214.846  &  1.9707  &  0.0051  & 754.685  &  1.9094  &  0.0298  \\
        226.909  &  1.9681  &  0.0056  & 800  &  1.9053  &  0.032  \\
        239.648  &  1.9655  &  0.0061  & 841.807  &  1.9018  &  0.0342  \\
        253.103  &  1.9629  &  0.0067  & 900  &  1.8947  &  0.037  \\
        267.314  &  1.9603  &  0.0073  & 1000  &  1.8807  &  0.0413  \\
        282.322  &  1.9576  &  0.0079  & 1100  &  1.8636  &  0.0457  \\
        298.173  &  1.9549  &  0.0086  & 1200  &  1.844  &  0.05  \\
        314.914  &  1.9522  &  0.0092  & 1300  &  1.8228  &  0.0544  \\
        332.594  &  1.9492  &  0.01    & 1400  &  1.7975  &  0.0588  \\
        \end{tabular}
    \label{Tab:free_energy_frenkel_pair}
\end{table}

\section{Normal Mode Decomposition}
\begin{figure}[h!]
    \centering
    \makebox[\columnwidth][c]{\includegraphics[width=1.\columnwidth]{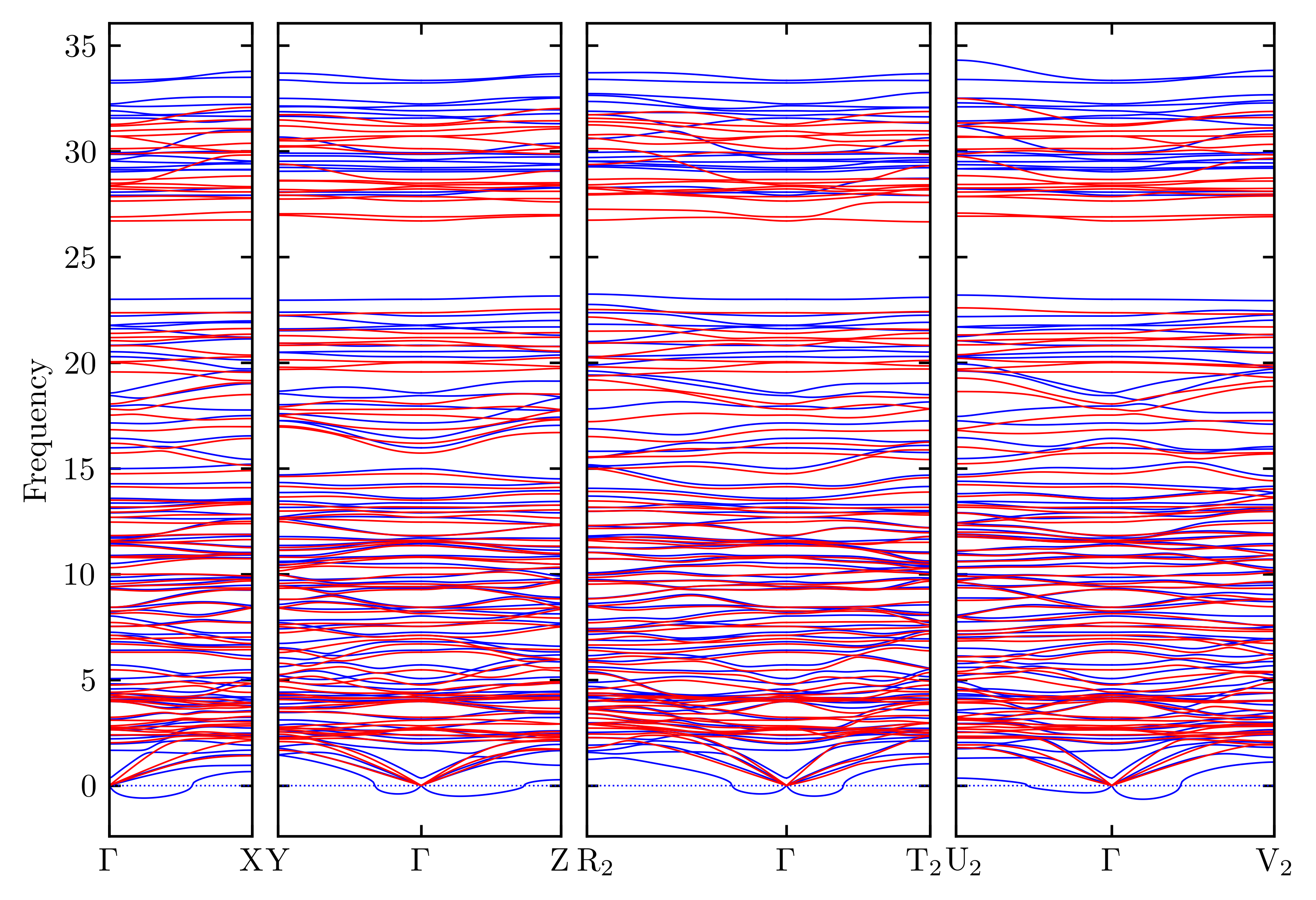}}
    \caption{The phonon bandstructure in the harmonic approximation (blue) and using normal-mode-decomposition (red) at the interpolated box dimensions corresponding to \SI{1325.60}{\kelvin}. The path is the one suggested in \cite{HINUMA2017140} for the reciprocal “reduced” cell having all-acute interaxial angles.
    }   \label{Fig:complete_bandstructure_comp}
\end{figure}

\begin{figure}[h!]
    \centering
    \makebox[\columnwidth][c]{\includegraphics[width=0.9\columnwidth]{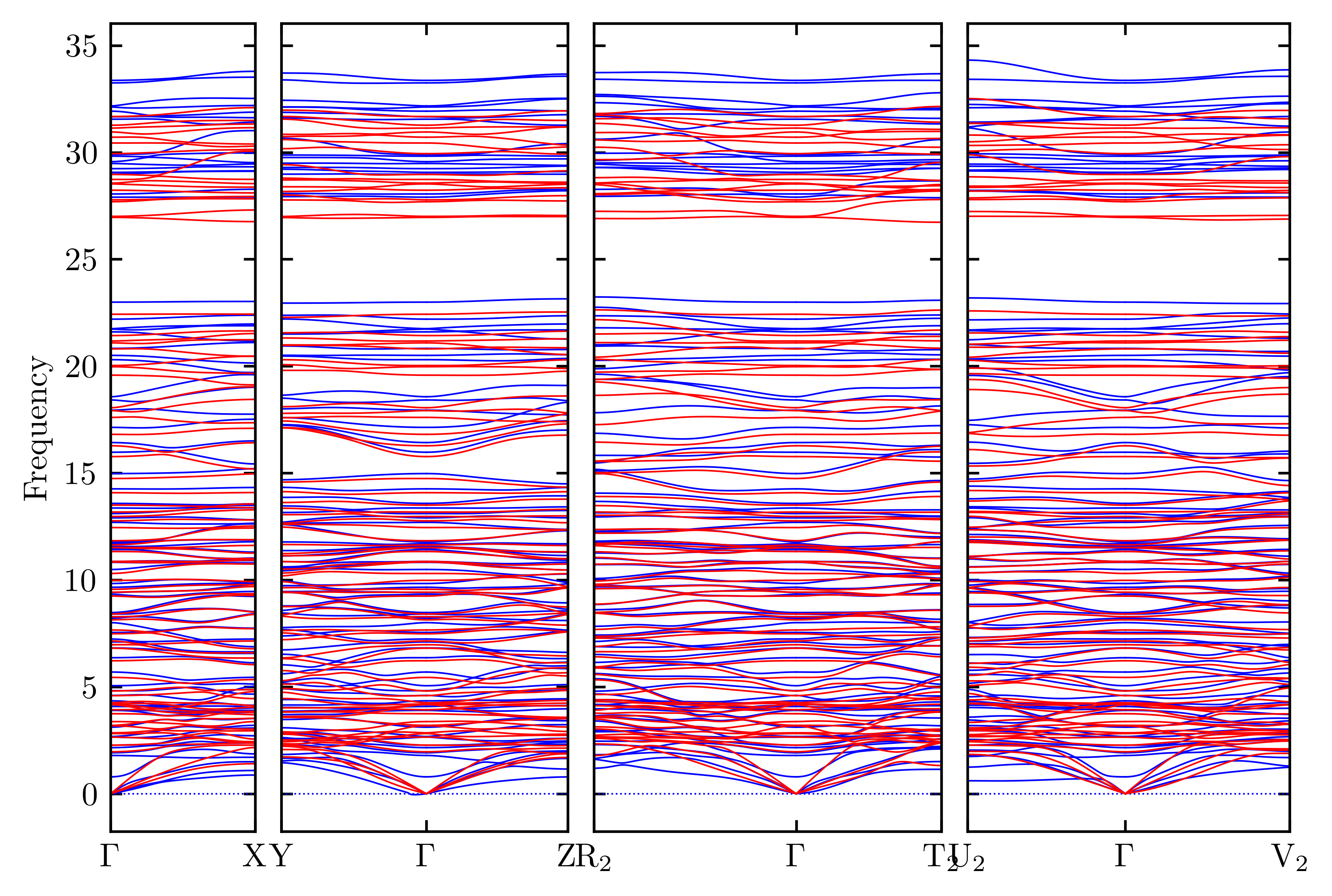}}
    \caption{\SI{1300}{\kelvin}.
    }   \label{Fig:combo1300}
\end{figure}

\begin{figure}[h!]
    \centering
    \makebox[\columnwidth][c]{\includegraphics[width=0.9\columnwidth]{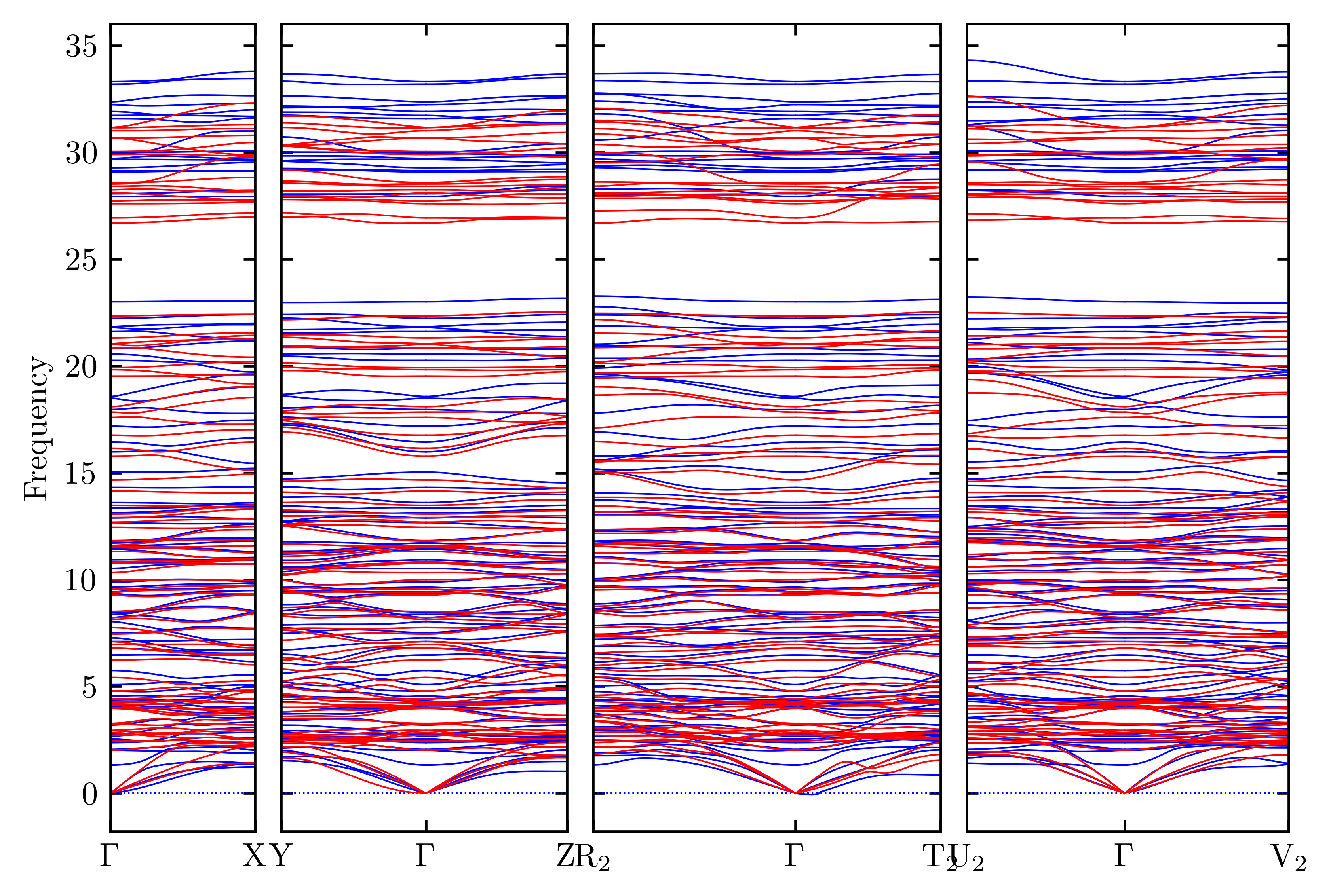}}
    \caption{\SI{1400}{\kelvin}.
    }   \label{Fig:combo1400}
\end{figure}

Anharmonic features of the phonon bandstructure for the pristine system were determined using the method of normal-mode decomposition \cite{Sun2014} implemented in dynaphopy \cite{CARRERAS2017221}. Supercells of size \begin{small}\(2\!\times\!2\!\times\!2\)\end{small} were first equilibrated in the NVT-ensemble for \SI{250}{\pico\second} using a Langevin-thermostat. Then they were equilibrated in the NVT-ensemble for another \SI{250}{\pico\second} using a Nose-Hoover thermostat. After which a trajectory was produced for \SI{3}{\nano\second} in the NVT-ensemble using the same Nose-Hoover thermostat. A timestep of \SI{1}{\femto\second} was used and configurations containing the velocities of all particles were saved each timestep.\\
Figure~\ref{Fig:complete_bandstructure_comp} shows the complete anharmonic bandstructure together with the harmonic approximation at the box dimensions showing a minimum in an optical mode. The imaginary frequencies in the harmonic approximation do not disappear for additional Q-points and the plotted harmonic bandstructure results from a 6$\times$6$\times$6 supercell.
The fact that the soft mode disappears below \SI{1300}{\kelvin} and if we extrapolate above \SI{1400}{\kelvin} in the harmonic approximation is evidenced in the bandstructures of these respective temperatures in Fig.~\ref{Fig:combo1300} and Fig.~\ref{Fig:combo1400}. The imaginary frequencies have almost disappeared in both cases.

\clearpage

%